\documentclass[aps,pre,twocolumn,superscriptaddress,showpacs]{revtex4-1}

\usepackage{amsmath,amssymb,bm}
\usepackage{graphics}
\usepackage{graphicx}

\def\er{Erd\H{o}s-R\'enyi }

\begin{document}

\title{Critical dynamics of the $k$-core pruning process}

\author{G. J. Baxter}
\affiliation{Department of Physics \& I3N, University of Aveiro, Campus
  Universit\'ario de Santiago, 3810-193 Aveiro, Portugal}
\author{S. N. Dorogovtsev}
\author{K.-E. Lee}
\affiliation{Department of Physics \& I3N, University of Aveiro, Campus
  Universit\'ario de Santiago, 3810-193 Aveiro, Portugal}
\author{J. F. F. Mendes}
\affiliation{Department of Physics \& I3N, University of Aveiro, Campus
  Universit\'ario de Santiago, 3810-193 Aveiro, Portugal}
\author{A. V. Goltsev}
\email[]{goltsev@ua.pt}
\affiliation{Department of Physics \& I3N, University of Aveiro, Campus
  Universit\'ario de Santiago, 3810-193 Aveiro, Portugal}
\affiliation{A. F. Ioffe Physico-Technical Institute, 194021
  St. Petersburg, Russia}


\begin{abstract}
We present the theory of the $k$-core pruning process (progressive removal of nodes with degree less than $k$) in uncorrelated
random networks. We derive exact equations describing this process and the evolution of the network structure, and solve them numerically and, in the critical regime of the process, analytically.
We show that the pruning process exhibits three different behaviors
depending on whether the mean degree $\langle q \rangle$ of the initial network is above, equal to, or below the threshold $\langle q \rangle_c$ corresponding to the emergence of the giant $k$-core. We find that above the threshold
the network relaxes exponentially to the $k$-core.
The system manifests the phenomenon known as ``critical slowing down'', as the relaxation time diverges when $\langle q \rangle$ tends to $\langle q \rangle_c$.
At the
threshold,
the dynamics become critical
characterized by a power-law relaxation ($\propto 1/t^2$).
Below the threshold,
a long-lasting transient process (a ``plateau'' stage) occurs.
This transient process ends with a collapse in which the entire network disappears completely. The duration of the process diverges when $\langle q \rangle \rightarrow \langle q \rangle_c$. We show that the critical dynamics of the pruning
are determined by branching processes of spreading damage.
Clusters of nodes of degree exactly $k$ are
the  evolving substrate for these branching processes.
Our theory
completely describes this branching cascade of damage in uncorrelated networks by providing the time dependent distribution function of branching.
These theoretical results are supported by our simulations of the $k$-core pruning in Erd\H{o}s--R\'enyi graphs.

\end{abstract}


\maketitle


\section{Introduction}

Pruning algorithms for networks provide an effective way to extract
subgraphs distinguished by their
structural properties, connectivity, robustness against failures and
damaging, and other features \cite{seidman1983network,
  bollobas1984, luczak1991size, batagelj2002generalized, dgm2006,
  Buldyrev2010}.
In general pruning processes, parts of a network are progressively
removed from it according to some rule. If the rule is simply random
removal of nodes, we obtain ordinary percolation \cite{ab2002,Newman2003,dgm2008}, but
in general we are interested in more complex pruning rules. The parts
removed may be
nodes \cite{seidman1983network,pittel1996,dgm2006}, clusters
\cite{bauer2001core}, finite connected components in interdependent
and multiplex networks
\cite{Buldyrev2010,zhou2012dynamics,zhou2014simultaneous,son2012percolation}, etc.
Despite the wide
variety of pruning processes, many of them demonstrate
similar behaviors, such as discontinuous hybrid phase transitions. The
$k$-core pruning as the simplest pruning process of this kind, stands
as a paradigm for all such pruning processes, so its theory should
help to understand the behavior of these pruning algorithms in general.
The $k$-core is the network subgraph
in which all nodes have degree
at least $k$ \cite{bollobas1984}. Since $k$-cores represent the densest parts of networks,
they play an important role in understanding the structure and dynamics of complex
network systems \cite{dgm2008}.
The
standard algorithm for finding the $k$-core of a network employs the following
pruning process: at each step remove all nodes of degree less than
$k$. This removal decreases the degrees of remaining
nodes, some of which will
become smaller than $k$. So, the pruning
is repeated until
either the $k$-core remains
or the network disappears
\cite{pittel1996}.

Previous investigations have mainly focused on the
final result of the $k$-core pruning process, namely the $k$-core.
These were the studies
which showed that $k$-core percolation is
a hybrid phase
transition, combining discontinuity and a critical singularity, in contrast to ordinary percolation (continuous phase transition)
\cite{dgm2006,dgm2008,dorogovtsev2006k}.
However, associating the number of steps in the pruning process
with time $t$ reveals a process exhibiting complex dynamics above,
below, and at the $k$-core percolation threshold.
Understanding
the
$k$-core pruning process and accompanying structural changes can shed
light on such physical phenomena as the jamming transition, the
rigidity percolation, and glassy dynamics
\cite{Birolli2007}. Furthermore, the $k$-core pruning process is one
of the simplest examples of dynamic processes associated with hybrid
phase transitions, sharing, for example, some common properties with cascade
failures in interdependent networks that have recently received significant
attention in the literature \cite{Buldyrev2010,zhou2013,Grassberger2015,zhou2012dynamics,zhou2014simultaneous,boccaletti2014structure,kivela2014multilayer,baxter2012avalanche}.

In this paper we develop the detailed theory
of the
$k$-core pruning process in uncorrelated, sparse random networks,
describing the temporal evolution of the networks structure,
the spreading of damage over the network, and critical phenomena in this process.
We show that near the threshold value of the mean degree, $\langle q \rangle_c$, corresponding to the emergence of the giant $k$-core, this cascade of removals of nodes is a branching process with the mean branching coefficient close to $1$. Our theory describes this
process completely providing the full time dependent distribution of branching from the beginning until the end of the pruning. We indicate that the clusters of nodes of degree $k$ (so-called ``corona clusters''), evolving due to the pruning, provide the substrate for the branching processes.
Near the threshold we find three different behaviors
depending on whether the mean degree $\langle q \rangle$ of the initial network is above, equal to, or below the threshold.
First, we demonstrate that above the threshold, the network relaxes exponentially to the steady $k$-core. The relaxation time diverges when $\langle q \rangle$ tends to $\langle q \rangle_c$, manifesting a phenomenon known as ``critical slowing down''. Second, at the critical point, $\langle q \rangle= \langle q \rangle_c$, the dynamics is critical,
characterizied by a power-law relaxation with $1/t^2$ dependence. Third, below the threshold, a long-lasting transient process (a ``plateau'' stage) occurs. This transient process ends with a collapse in which the entire network disappears. We find that the duration of the process diverges when $\langle q \rangle$ approaches  $\langle q \rangle_c$. Our theory is supported by numerical calculations for Erd\H{o}s--R\'enyi graphs and by simulations of the pruning process in these random graphs.

In Sec.~\ref{equations} we derive the exact equations describing the
evolution of the network structure during the pruning process enabling
us to obtain the time dependent degree distribution $P(q,t)$ and the
branching probability distribution ${\cal P}(n,t)$ at all times.
Close to the critical point,
the probability that different branches of the process cross each
other is negiligibly small.
We show that in this region, our equations take a simple form for analytical treatment.
Section~\ref{three regimes} explores the three regimes of the pruning
process below, at, and above the threshold. Section~\ref{branching
  processes} describes the statistics of the branching process.
A relationship with dynamical systems close to a saddle-point bifurcation and details of calculations are given in the appendices.


\section{Evolution equations}
\label{equations}

To study the $k$-core pruning process, let us consider as a
representative case an infinite
uncorrelated sparse random network, which is completely defined by its degree
distribution $P(q)$.
In this case, we can write exact equations for the evolution of the
degree distribution.
Let $P(q,t)$ be the proportion of vertices having degree $q$ at time
$t$, with the initial condition $P(q,0) = P(q)$. At each time $t =
1,2,3,...$, all vertices with degree $q$ less than $k$ are pruned by
having all edges connected to them removed from the network. The
probability $P(0,t)$ thus tracks the number of vertices pruned so far.


\subsection{Exact evolution equations}\label{exact equations}

The removal of edges from pruned vertices means that some non-pruned
vertices will also lose edges, changing the degree distribution of the
remaining network.
Let $r_t$ be the probability that, upon following a
randomly chosen edge within the network existing at time $t$, we
arrive at a vertex with degree less than $k$:
\begin{equation}
\label{r}
r_t = \frac{1}{\langle q\rangle_t}\sum_{q<k} q P(q,t)
.
\end{equation}
Such an edge will be removed in the subsequent step.
Here $\langle q\rangle_t$ is the mean degree of the surviving network
at time $t$,
\begin{equation}
\langle q\rangle_t = \sum_q q P(q,t)
.
\end{equation}
The
probability that a vertex of degree $q' \geq k$ at time $t$ has $q$ surviving
edges at time $t+1$ is then
$\binom{q'}{q}(1{-}r_t)^{q}r_t^{q'-q}$. A vertex of degree $q' < k$ at
time $t$ will of course have degree zero at time $t+1$. Summing over
all $q'$, the degree distribution then evolves as follows:
\begin{equation}
\label{Pupdate}
P(q,t+1) = \sum_{q' \geq \max\{q,k\}} P(q',t) \binom{q'}{q}  (1-r_t)^{q} r_t^{q'-q}
\end{equation}
for $q > 0$ while the fraction of pruned nodes evolves according to
\begin{equation}
\label{P0update}
P(0,t+1) = \sum_{q' < k} P(q',t)
,
\end{equation}
where the sum includes $q' = 0$.
The uncorrelated nature of the network ensures that
Eqs.~(\ref{r})--(\ref{P0update}) completely define the evolution of the
network at all times.
Note that another approach for the pruning process which, however, does not consider the evolution of  the network structure, was proposed in \cite{iw2009}.

\begin{figure*}[htb]
\includegraphics[width=1.2\columnwidth]{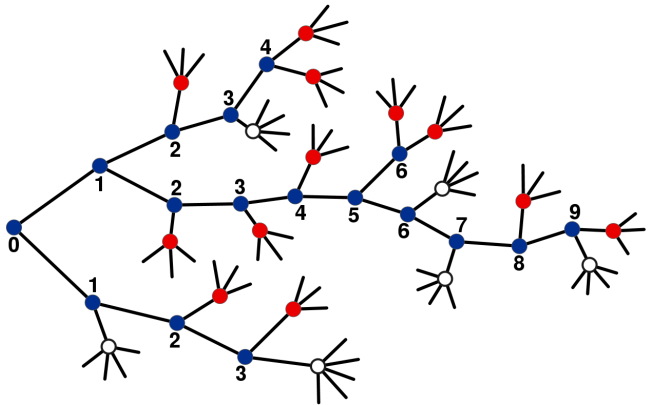}
\caption{(Color online) A snapshot of
the branching process of propagation of node pruning in a small part of the network of $10^5$ nodes
during the plateau stage ($\langle q \rangle < \langle q \rangle_{c}$) of the $k$-core pruning process for $k=3$.
  The node labelled $0$ is pruned, causing the
  corona nodes (i.e., nodes with degree $3$) labelled $1$ to lose
  edges. These two nodes are pruned in the next step, and so on, with
  further corona nodes removed in subsequent steps according to the
  numbered order. Red
    and white circles represent the nearest neighboring nodes
    of degree $4$ and greater than $4$, respectively, that
    survive because their degrees exceed $3$. The red nodes
    after this pruning become of degree $3$. They  augment other
    corona clusters, which may then be pruned at a later time.}
\label{fig:pruning_cluster}
\end{figure*}

To understand the spreading of damage through the network as the
pruning process evolves,
we introduce the probability $s_t$.
This is the probability that, following
an edge at time $t$, we reach a vertex that has degree at least $k$ at
time $t$, but will have no more than $k-1$ other surviving edges at time
$t+1$ (not counting the edge through which we reached the
vertex). This means that if the edge we are
following is removed at time $t$, the vertex that it leads to will be
removed at time $t+1$. To calculate $s_t$, we sum over probabilities
that all but $l$ of the $q-1$ outgoing edges of a vertex of degree $q$
(i.e. $q-1-l$ edges)
are lost at time $t$ (each one with probability $r_t$) with $l$ equal
to at most $k-1$. A second summation is then performed over all possible
degrees $q \geq k$:
\begin{equation}
\label{s_exact}
s_{t} = \frac{1}{\langle q\rangle_t}\sum_{q \geq k}q P(q,t)
 \sum_{l = 0}^{k-1} \binom{q-1}{l}r_t^{q-1-l}(1-r_t)^{l}.
\end{equation}
The probability $\mathcal{P}(n,t)$ that a vertex removed at time
$t$ has $n$ neighbors that will be removed at time $t+1$ is then
\begin{equation}\label{branching_dist_exact}
\mathcal{P}(n,t) = \frac{\sum_{q=n}^{k-1} P(q,t)
\binom{q}{n} s_t^n(1-s_t)^{q-n}}{\sum_{q=1}^{k-1} P(q,t)}
.
\end{equation}
This function describes the branching of spreading damage. The mean branching is
\begin{equation}
b_t = \sum_{n = 0}^{k-1} n \mathcal{P}(n,t)
 = s_t \frac{\sum_{q=1}^{k-1}q
  P(q,t)}{\sum_{q=1}^{k-1} P(q,t)}
  .
  \label{mean_branching exact}
\end{equation}


\subsection{Non-crossing approximation}\label{non-crossing appr}

Unfortunately, it is difficult to study analytically Eqs.~(\ref{r})--(\ref{P0update}). In this subsection we develop an approximate approach
providing the asymptotic description of the pruning process at large times near the critical point.

When the probability $r_t$ is very small, the pruning can then be considered as a branching
process. The probability that a vertex loses two neighbors in a single step is
negligible, in other words, the probability that two or more branching
trees meet at a vertex is negligible. The process then evolves with independent branching trees spreading simultaneously
over the network.
An example of such non-crossing branchings observed in simulations
is shown in Fig.~\ref{fig:pruning_cluster}.
If crossings are negligible, then the fraction of vertices of degree $q < k-1$ is also negligible and only vertices of degree $q\geq k-1$ must be taken into account. This is the main assumption of the  ``non-crossing'' approximation.
This approximation is supported by our numerical solution of Eqs.~(\ref{r})--(\ref{P0update}) and simulations which show that the probability of crossings between branches are negligible and $P(k-1,t) \gg P(k-2,t) \gg \dots P(1,t)$ already after a short initial period (see the next sections).
Applying the non-crossing approximation to Eq.~(\ref{s_exact}), we find
that $s_t$ becomes simply the probability that,
 following an edge at time $t$, we encounter a vertex of degree $k$.

\begin{equation}
\label{s}
s_{t} \approx \frac{k P(k,t)}{\langle q\rangle_t}
.
\end{equation}
Furthermore, the probability $r_t$, Eq.~(\ref{r}), and the mean
branching $b_t$, Eq.~(\ref{mean_branching exact}), take the simple forms,
\begin{eqnarray}
&&
r_t \approx \frac{ (k-1) P(k-1,t)}{ \langle q\rangle_t}
,
\label{r approx}
\\[5pt]
&&
b_t \approx \frac{ (k-1)k P(k,t)}{ \langle q\rangle_t}
.
\label{mean_branching}
\end{eqnarray}
So $r_t$ is simply the probability that, following an edge at time $t$, we encounter a vertex of degree $k-1$.
The evolution equation (\ref{Pupdate}) is also simplified. The following set of equations determines the evolution of the degree distribution during the $k$-core pruning process:
\begin{eqnarray}
&&\!\!\!\!\!\!\!\!
P(q,t{+}1) {=} P(q,t){-}r_t q P(q,t){+}r_t (q{+}1)P(q{+}1,t),
\label{q-generation}
\\[5pt]
&&\!\!\!\!\!\!\!\!
P(k-1,t+1)= r_t k P(k,t),
\label{k-1-generation-1}
\\[5pt]
&&\!\!\!\!\!\!\!\!
P(0,t+1) = P(0,t)+P(k-1,t),
\label{0-generation-1}
\\[5pt]
&&\!\!\!\!\!\!\!\!
\langle q\rangle_t =(k-1)P(k-1,t)+\sum_{q\geq k} qP(q,t),
\label{mean-degree-appr}
\end{eqnarray}
where $q\geq k$.
The negative term in Eq. (\ref{q-generation}) corresponds to the reduction
in $P(q,t)$ due to vertices of degree $q$ losing with the probability $q r_t$ a single
edge, while the positive term (last term) corresponds to an
increase in $P(q,t)$ due to vertices of degree $q+1$ losing
an edge with the probability $(q+1)r_t$  and so ending up with degree $q$.

Using Eq.~(\ref{mean_branching}), we rewrite Eq.~(\ref{k-1-generation-1}) as follows,
\begin{equation}
\label{k-1-generation}
P(k-1,t+1)= b_t P(k-1,t)
.
\end{equation}
Equations~(\ref{k-1-generation-1}) and (\ref{k-1-generation}) show that
the removal of a vertex of degree $k{-}1$ at time $t$ triggers in the next step
the removal of all corona vertices attached to it since they will lose one edge and will have degree $k{-}1$. On average, the number of these corona vertices is the mean branching $b_t$.
In uncorrelated networks, Eqs.~(\ref{q-generation})--(\ref{mean-degree-appr})
describe the non-crossing branching processes of spreading damage (see Appendices \ref{crit-relax-appr}  and \ref{plateau-appr}).
They show that vertices of degree $k$ (``corona'' vertices) are crucial for spreading damage. In the case $\langle q\rangle \geq \langle q\rangle_c$ at large times, $t \gg 1$, crossings are negligible and these equations are asymptotically exact.
Equations~(\ref{q-generation})--(\ref{mean-degree-appr}) are not valid when there are numerous crossings between branching processes. Such crossings are abundant both at the initial stage of the pruning process and at the end of the ``plateau'' stage when the network collapses.
In this case, the exact Eqs.(\ref{r})--(\ref{P0update}) must be used.
Branching processes are discussed in detail below in
Sec. \ref{branching processes}.


\section{Three regimes of the pruning process}
\label{three regimes}

As a representative example of the pruning process, we solved
Eqs.~(\ref{r})--(\ref{P0update}) numerically for \er networks
(Poisson degree distributions) using the initial mean degree
$\langle q \rangle$ as a control parameter. We solved the equations for
$k=3$ and $k=5$.
The $3$-core appears with a hybrid
transition at $\langle q\rangle_c \approx 3.35091887$, while for the $5$-core,
$\langle q \rangle_c \approx 6.7992755$. We also performed simulations of the pruning process in the networks. We found that for any $k \geq 3$, the dynamics of the pruning process can be divided into three different regimes: $\langle q \rangle < \langle q \rangle_c$, $\langle q \rangle > \langle q \rangle_c$, and $\langle q \rangle = \langle q \rangle_c$.


\subsection{Pruning process below $\langle q \rangle_c$}
\label{plateau-stage}

\begin{figure}[htb]
\includegraphics[width=1\columnwidth]{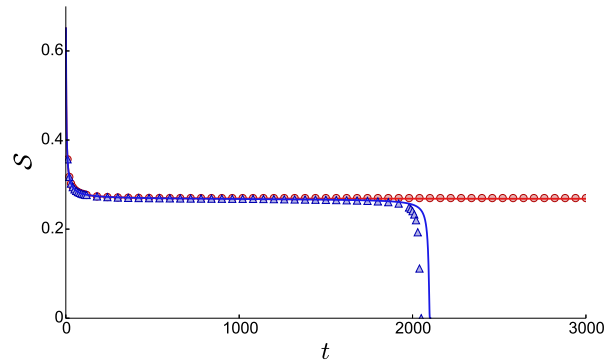}
\caption{(Color online) Size $\mathcal{S}$ of the \er network vs time
  $t$ during the pruning process for $k=3$ in two cases: (i)  below the
  threshold $\langle q \rangle_c$,  the system passes through a long
  ``plateau'' stage before a final collapse. Shown are numerical
  calculations for mean degree $\langle q \rangle =
  3.3509$ (blue solid line) and simulations (triangles) for a network
  of $10^8$ vertices showing similar total time.
(ii) Above $\langle q \rangle_c$, the system relaxes to a finite size,
 numerical solution for  $\langle q \rangle = 3.35092$ (red solid
 line) and simulations (circles).
\label{fig:plateau}}
\end{figure}

Below $\langle q \rangle_c$, the pruning process ends in a finite time (number of steps)
with the complete destruction of the infinite network.
Rapid pruning of vertices at early times soon
slows down and the system enters a ``plateau'' stage in which the rate of
removal of vertices is very slow. Finally, this transient process ends
with a collapse in which the entire network disappears, as can be seen
in Fig.~\ref{fig:plateau} which displays the temporal dependence of the network size  $\mathcal{S}$.
The duration of the entire process, from beginning until final collapse, diverges as the inverse square root of the
distance from the critical point,
\begin{equation}
\label{T_divergence}
T =A_{\text{below}}/\sqrt{\langle q  \rangle_c-\langle q \rangle}
,
\end{equation}
as shown in Fig.~\ref{fig:times}. The time $T$ is mainly determined by the duration of the ``plateau'' stage. Note that the inverse square-root scaling law is a
general feature of non-linear dynamic systems that are close to a saddle-node bifurcation \cite{strogatz1989,Strogatz1994}. In such systems, the long-lasting transient process is caused by a bottleneck region (the ghost) that exists in phase space when the system is close to a saddle-node bifurcation or the limiting point of metastable states in the case of the first order phase transitions (see a simple model in Appendix~\ref{appendix}). The nature of the bottleneck effect in the
$k$-core pruning process is discussed in Sec.~\ref{branching processes}.

\subsection{Pruning process above $\langle q \rangle_c$}
\label{pruning above}

\begin{figure}[htb]
\includegraphics[width=1\columnwidth]{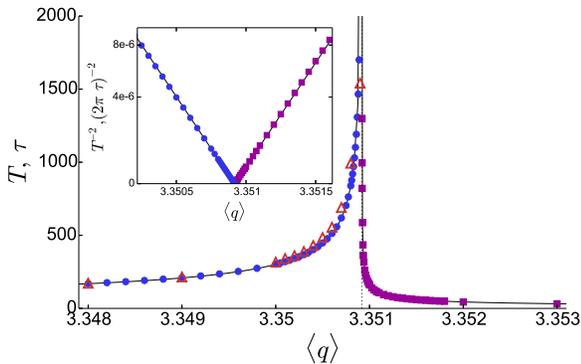}
\caption{(Color online) Characteristic times associated with the
  $k$-core pruning process for $k=3$ on an \er network. Circles
  show the duration $T$ of the entire pruning process below $\langle q \rangle_c$. Squares show the relaxation time constant $\tau$ above
  $\langle q \rangle_c$. Fitted square root scaling for $T$
  and $\tau$ are shown by black solid lines. The critical point
  $\langle q\rangle_c$ is marked by a vertical dotted line.
Completion times for a simulated network of $10^8$ vertices are also
shown below $\langle q \rangle_c$ (triangles).
 In the inset, the inverse squares of $T$ and $\tau$ (also scaled by
 $2\pi$) are shown, demonstrating the inverse square dependence on the
 distance from the critical point.
\label{fig:times}}
\end{figure}

Above $\langle q \rangle_c$ a finite fraction of
the network remains indefinitely and the network relaxes to the steady $k$-core only in the infinite time limit (see Fig.~\ref{fig:plateau}). In this regime, according to the numerical solution of Eqs.~(\ref{r})--(\ref{P0update}) and simulations, the relaxation to the
steady state is exponential. Instead of
measuring the total time, we
characterize the time scale of the pruning process by measuring the
relaxation time $\tau$, where:
\begin{equation}
\label{relaxation above}
P(k-1,t) \propto e^{-t/\tau}
.
\end{equation}
The relaxation time $\tau$ diverges as the inverse square root of
the distance from the critical point, as seen in Fig.~\ref{fig:times},
\begin{equation}
\label{tau_divergence}
\tau =A_{\text{above}}/\sqrt{\langle q
  \rangle_c-\langle q \rangle}.
\end{equation}
We examine the origin of this scaling in more detail in the next
Sec.~\ref{braching-above}, using the non-crossing approximation.

The divergence of $\tau$ manifests the phenomenon known as critical slowing down. Furthermore, comparing the amplitudes $A_{\text{below}}$ and $A_{\text{above}}$ of
the square-root singularities below and above the transition, we find
their ratio to be $A_{\text{below}}/A_{\text{above}} = 9.133/1.452 = 6.29$ for
$k=3$ and $8.44/1.34 = 6.28$ for $k=5$, in agreement with the ratio
$2\pi$ expected for general transitions of this kind, see Appendix~\ref{appendix}.

\begin{figure*}[htb]
\includegraphics[width=0.4\textwidth]{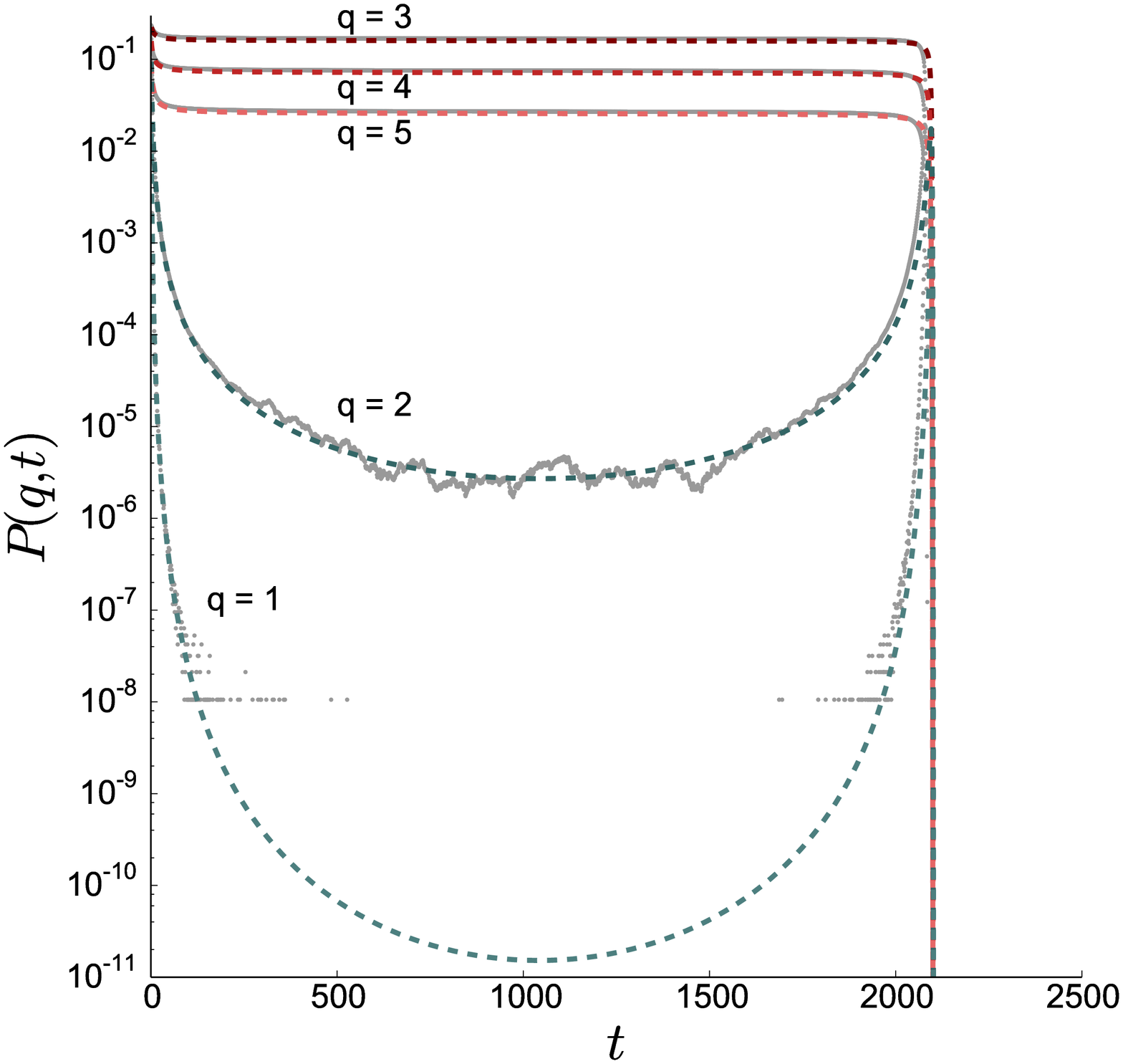}
\includegraphics[width=0.4\textwidth]{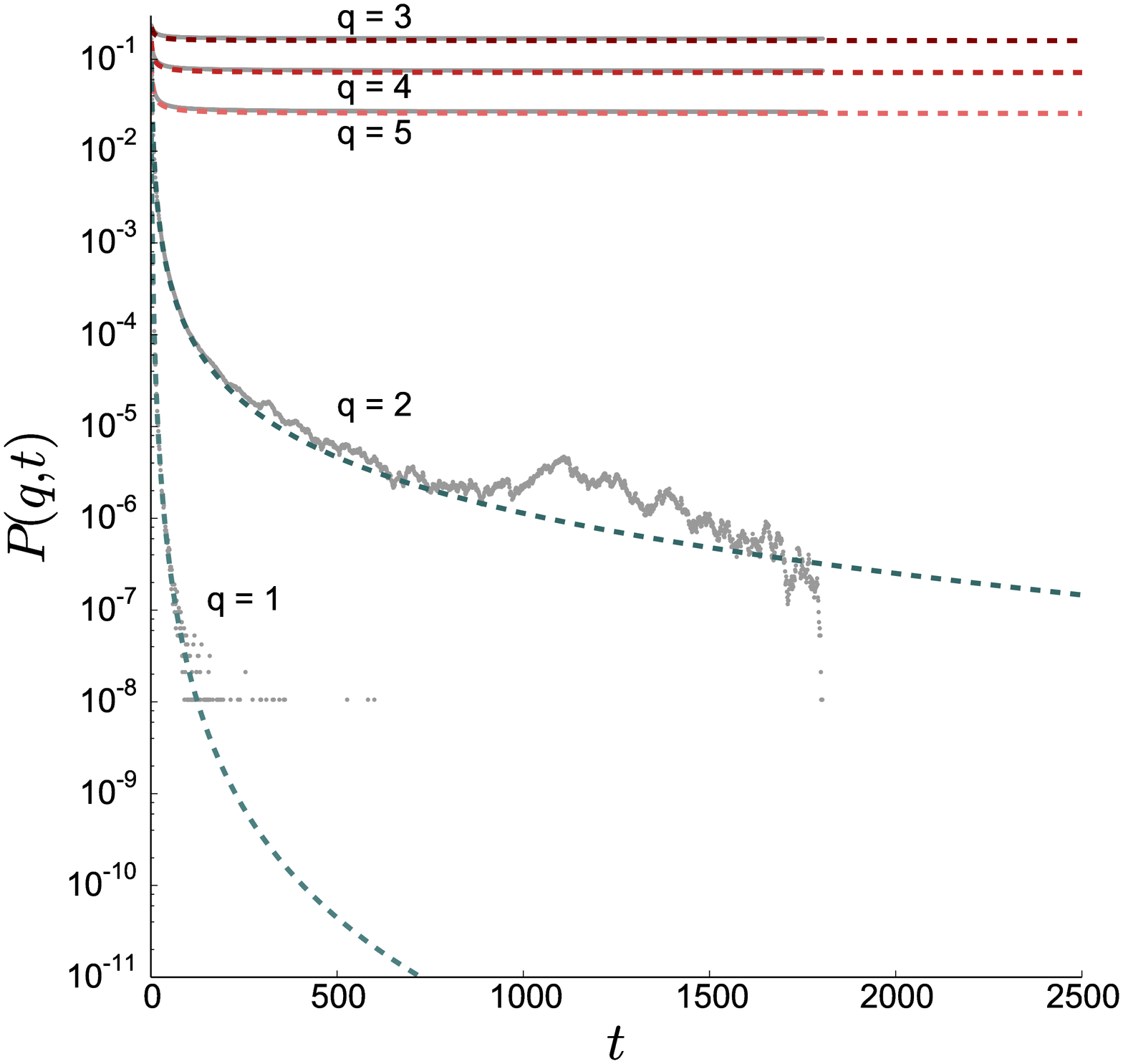}
\caption{(Color online) Time evolution of the network degree distribution during the
  $k$-core pruning process, for an \er network with $k=3$. Each line
  shows $P(q,t)$ for a different value of $q$, in order from top to
  bottom, $q = 3,4,5,2,1$, as labelled.
  (left) Initial mean degree $\langle q \rangle = 3.3509$, (right)
  initial mean degree $\langle q \rangle = 3.35092$.
Also shown are traces from simulation runs for an \er network with $N
= 10^8$ vertices and mean degree $3.3511$ (left) and $3.35111$
(right). Note that the critical point for a particular realisation is
a stochastic quantity,
so the mean degrees for matching theory and simulation are not
necessarily equal. Theoretical curves were chosen to be near the
critical point and to have a similar total time.
\label{fig:evolution}}
\end{figure*}

\begin{figure}[htb]
\includegraphics[width=1\columnwidth]{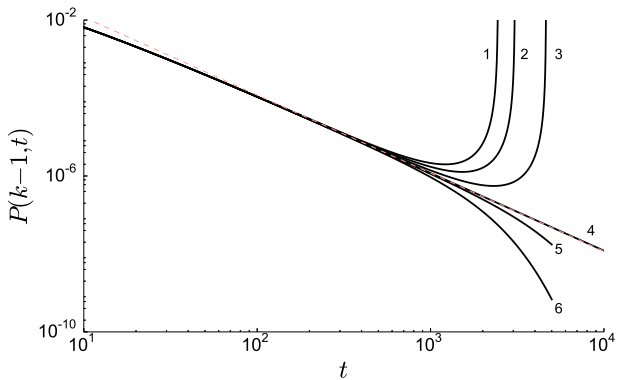}
\caption{Decay of $P(k-1)$ on a log-log scale for several values of
  $\langle q \rangle$
  close to $\langle q \rangle_c$. Exactly at the critical point, the
  decay follows a power law with exponent $-2$. (For clarity, the final
  collapse of $P(k-1)$ for curves below the critical point is not
  shown.) Curves are plotted for the Erd\H{o}s--R\'enyi graphs with
  the mean degree values (labelled $1$ to $6$)
 $3.350905$, $3.35091$, $3.350915$, which are below $q_c$,
$3.35091887$ (very close to $q_c$),
and $3.35092$, $3.350925$, which are above $q_c$.
Dashed line is a power-law decay with exponent $-2$.
\label{fig:powerlaw}}
\end{figure}

In Fig.~\ref{fig:evolution} we show the evolution of the degree
distribution just above and just below $\langle q \rangle_c$.
Near the critical mean degree the initial
evolution of the degree distribution $P(q,t)$ both above and below the
critical point is similar, namely, there is a sharp initial decrease of
$P(q,t)$ for nonzero $q$.
Below $\langle q \rangle_c$, however, the network finally collapses
completely, while above the critical point, the $k$-core survives forever.
The theoretical results agree
well with simulation.


\subsection{Critical pruning process}
\label{crit-prunning}

Solving Eqs.~(\ref{r})--(\ref{P0update}) numerically for \er networks, we find that exactly at the critical point, 
$\langle q \rangle_c$, the relaxation
is much slower, with $P(k-1,t)$ decaying as a power law,
\begin{equation}\label{critpowerlaw}
P(k-1,t) \propto \frac{1}{t^\sigma}.
\end{equation}
as can be seen in Fig.~\ref{fig:powerlaw}.
For $k=3$ we measured the exponent $\sigma = -1.993$ at $\langle q
\rangle = 3.35091887$, suggesting that the exponent is $-2$.
Note that
in
a simple model of a particle moving in a one-dimensional potential in Appendix~\ref{appendix}
the corresponding critical exponent is $-1$ [see Eq.~(\ref{a3})].
We explain the power-law behavior, Eq.~(\ref{critpowerlaw}), in Appendix~\ref{crit-relax-appr} by solving Eqs.~(\ref{B1})--(\ref{B3}) within the non-crossing approximation. This approach  gives the exact value $\sigma=2$.

\begin{figure*}[htb]
\includegraphics[width=0.8\textwidth]{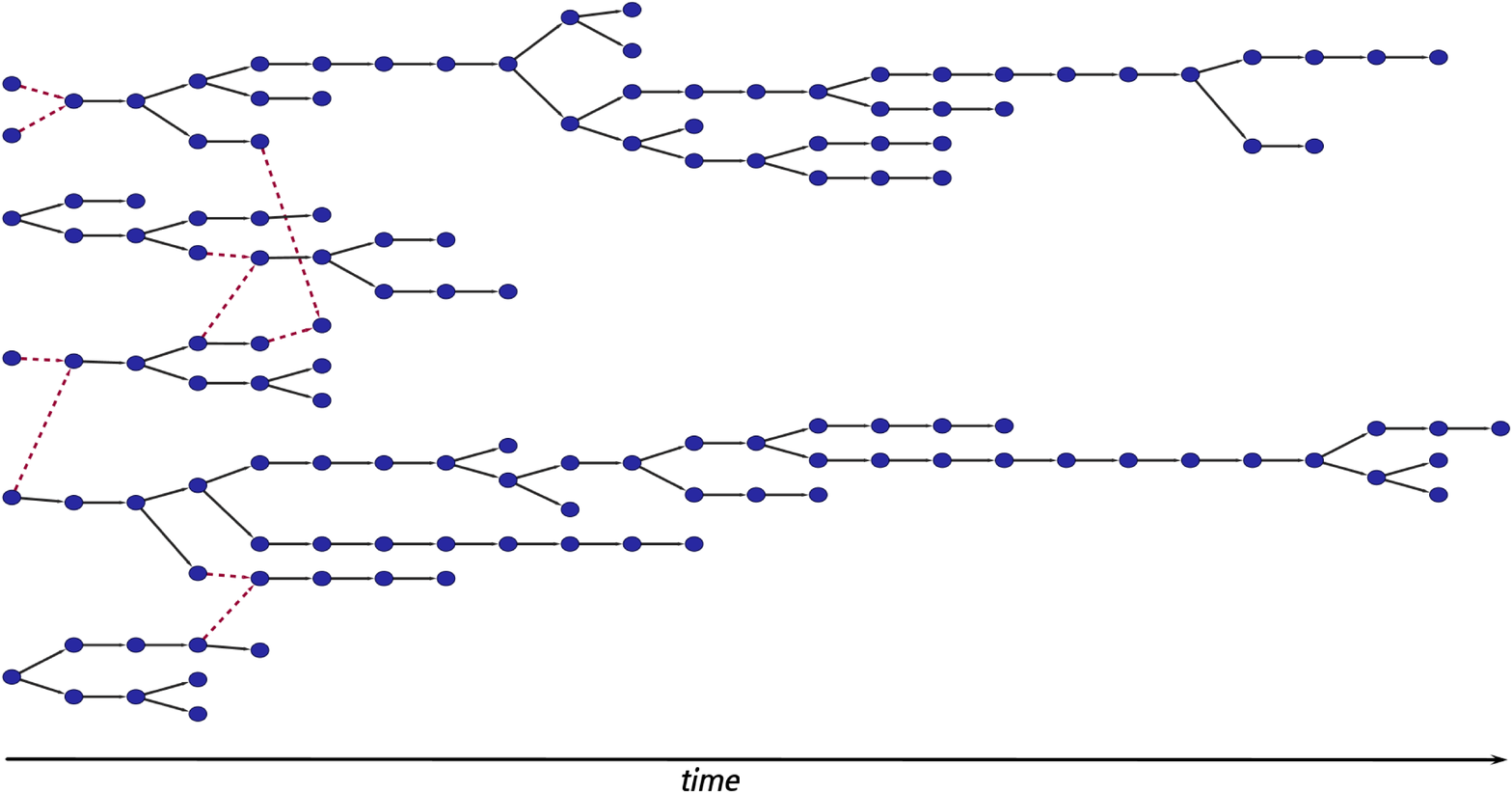}
\caption{(Color online) Example of the pruning process from time $t=10$ till 34 in a small part of the \er network with $10^5$ nodes. Time progresses from left to right in the tree. Blue circles represent vertices removed at a given time step. Their removal results in the removal of vertices on the right, and so on. Crossings (dashed lines) between the branching processes are abundant at the beginning of the pruning process.  They appear rarely after a short initial period.
Typical trees of medium size are shown. Much longer and much shorter trees also occur. }
\label{fig:branching_trees}
\end{figure*}

\begin{figure*}[htb]
\includegraphics[width=0.9\columnwidth]{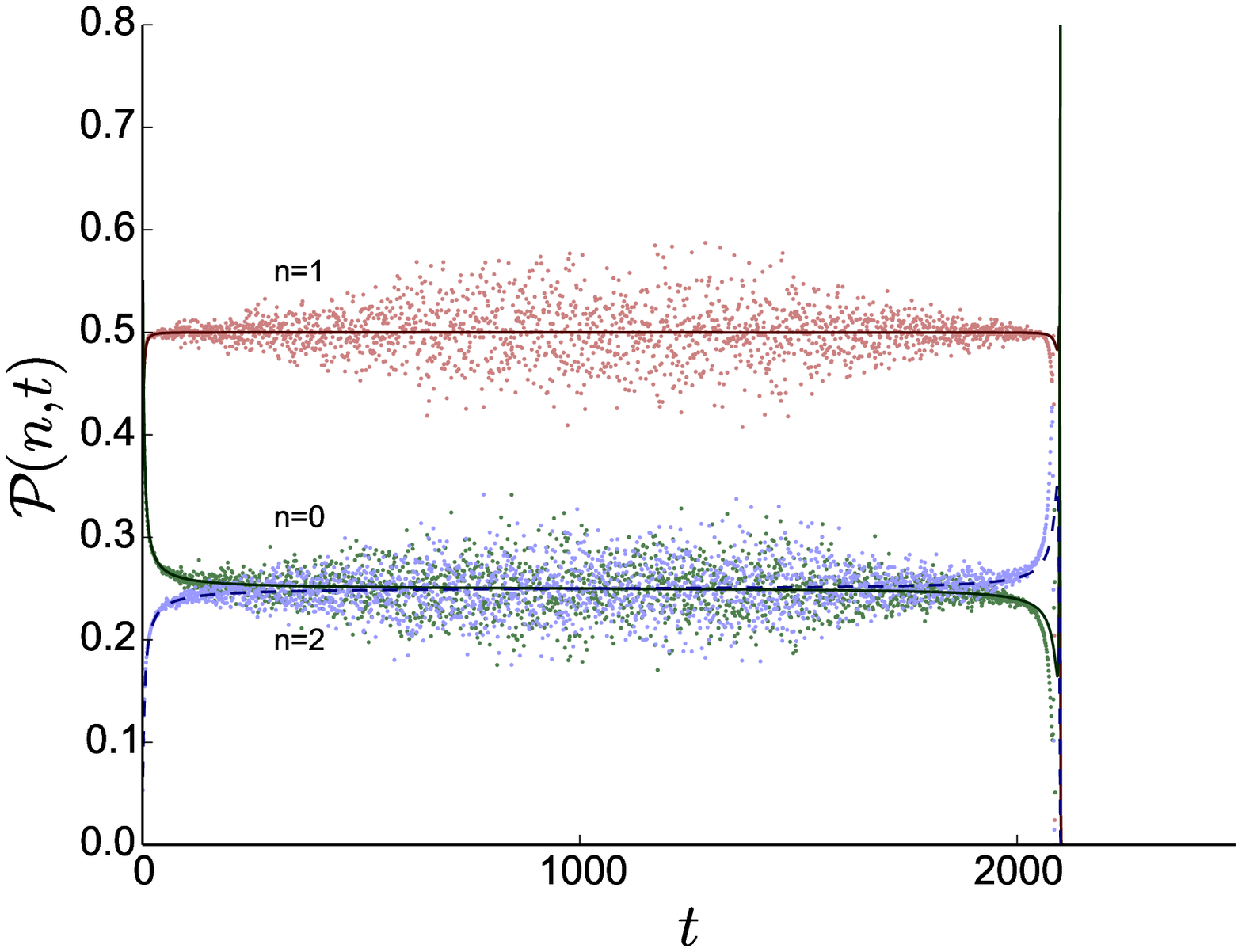}
\includegraphics[width=0.9\columnwidth]{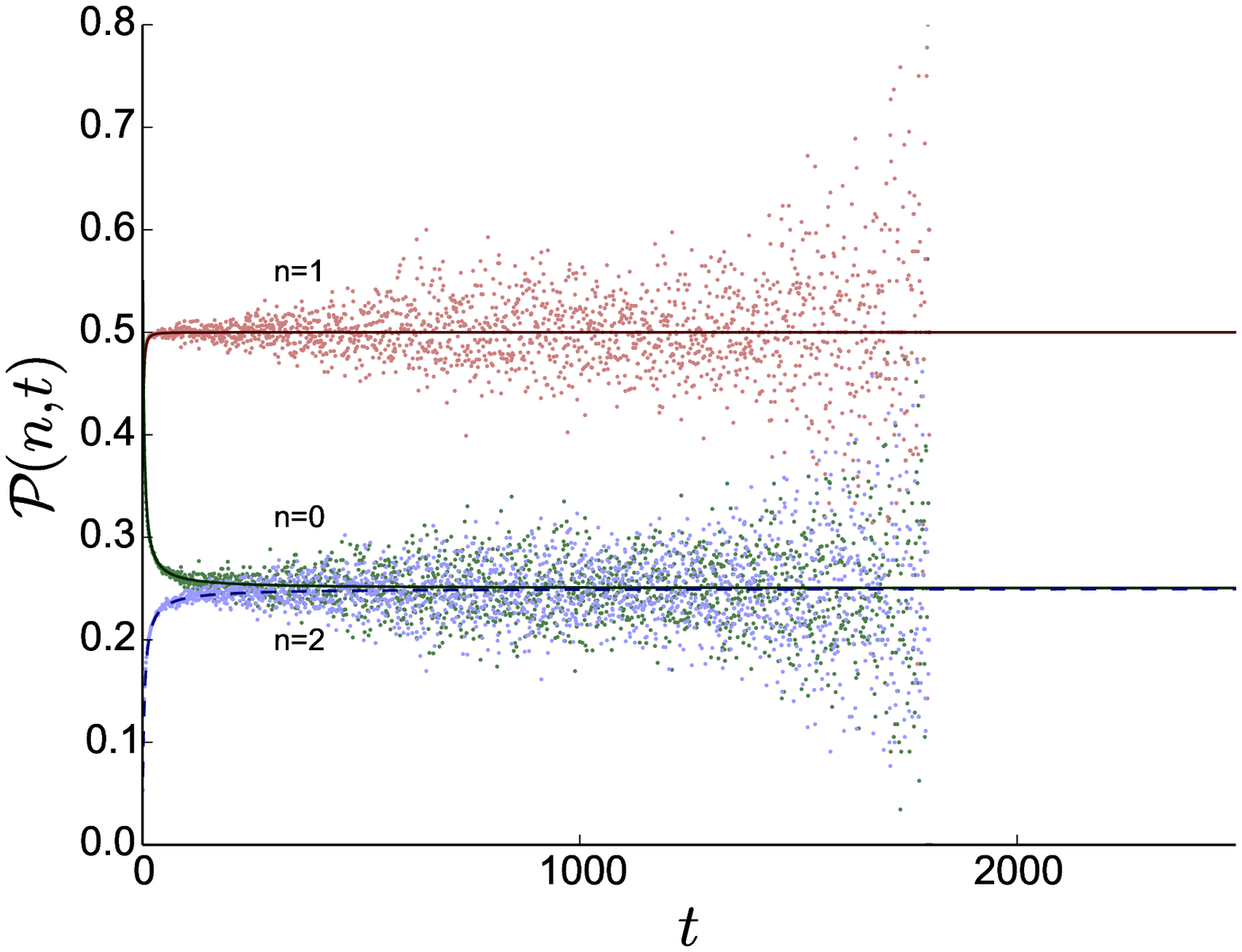}
\caption{(Color online) Evolution of the branching distribution $\mathcal{P}(n,t)$
  below (left) and above (right) the critical point for $k=3$. Solid
  and dashed curves are theoretical curves from
  Eq.~(\ref{branching_dist_exact}) for $n$ as labelled, for $\langle q
  \rangle = 3.3509$ (left) and $\langle q
  \rangle = 3.35092$ (right). Points are measured from simulation of
  an \er network of $10^8$ nodes, at $\langle q
  \rangle = 3.3511$ (left) and $3.35111$ (right), being just below and
  just above the critical mean degree for that network.
}\label{fig:branching_dist}
\end{figure*}

\begin{figure*}[htb]
\includegraphics[width=0.9\columnwidth]{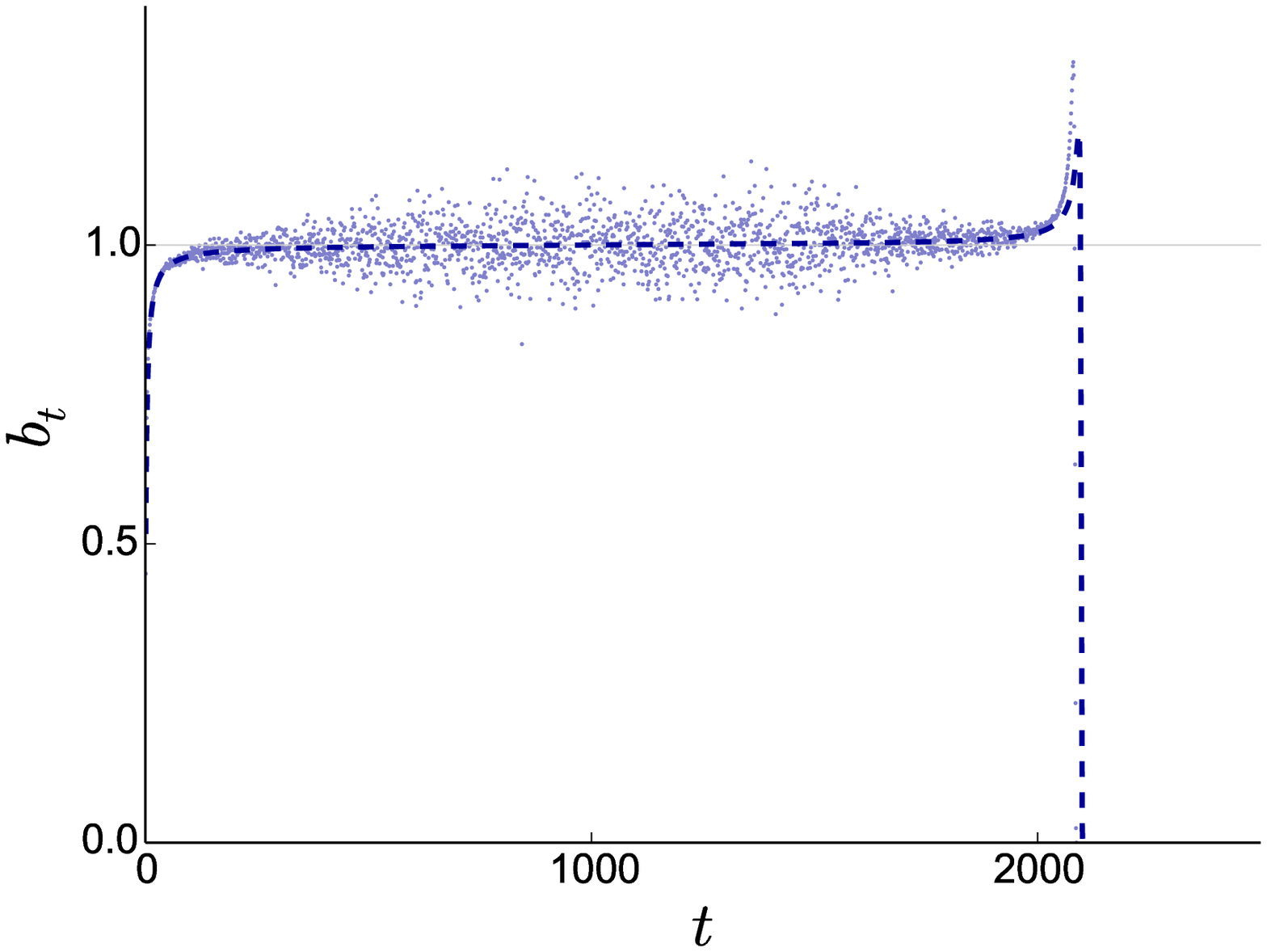}
\includegraphics[width=0.9\columnwidth]{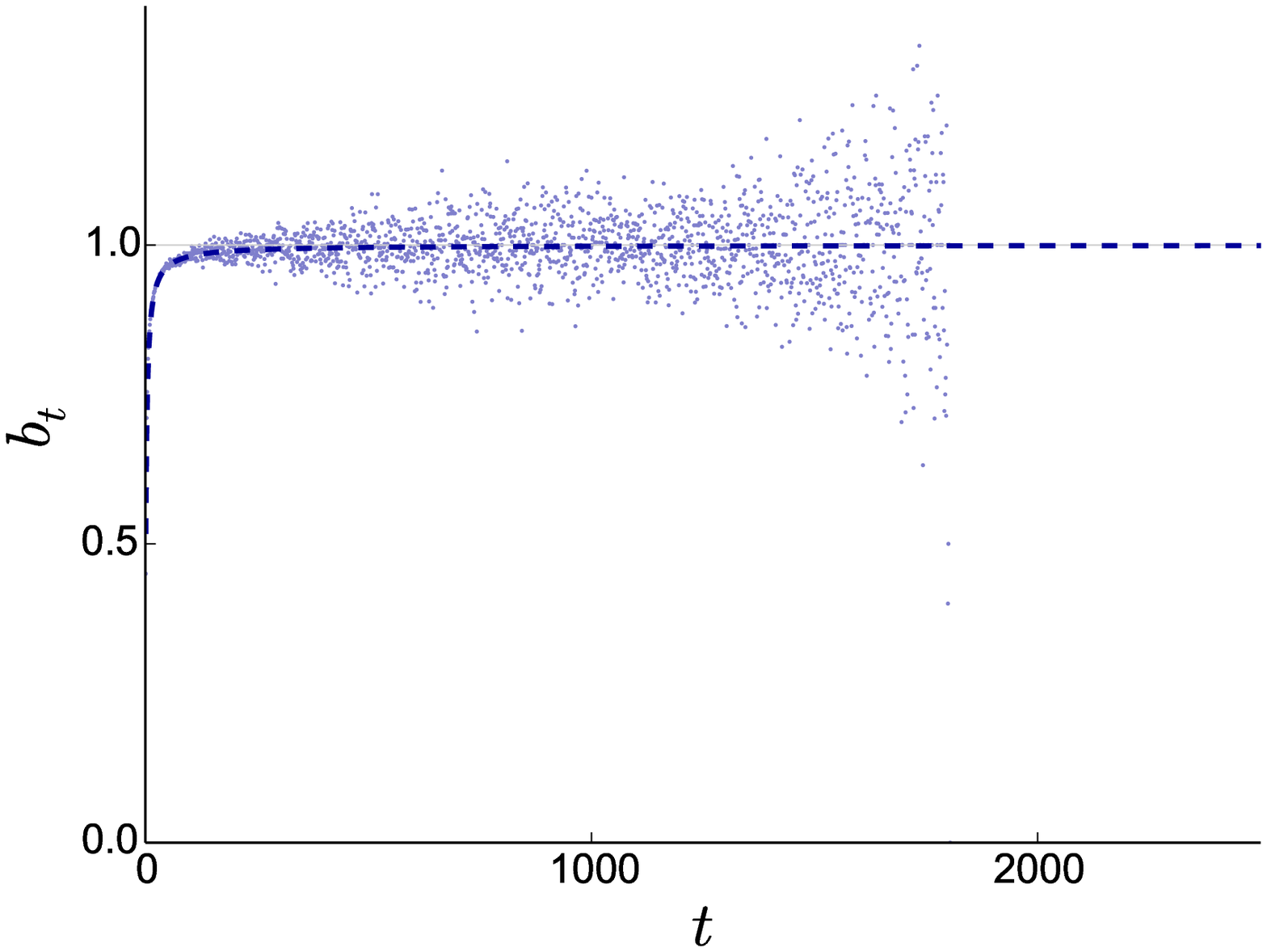}
\caption{(Color online) Evolution of the mean branching ratio $b_t$ below (left) and
  above (right) the critical point for $k=3$. Dashed curves are
  calculated using Eq.~(\ref{mean_branching}), points are from simulations.
Parameters used and simulation realisations are the same as in
Fig.~\ref{fig:branching_dist}.}
\label{fig:mean_branching}
\end{figure*}

\begin{figure}[htb]
\includegraphics[width=1\columnwidth]{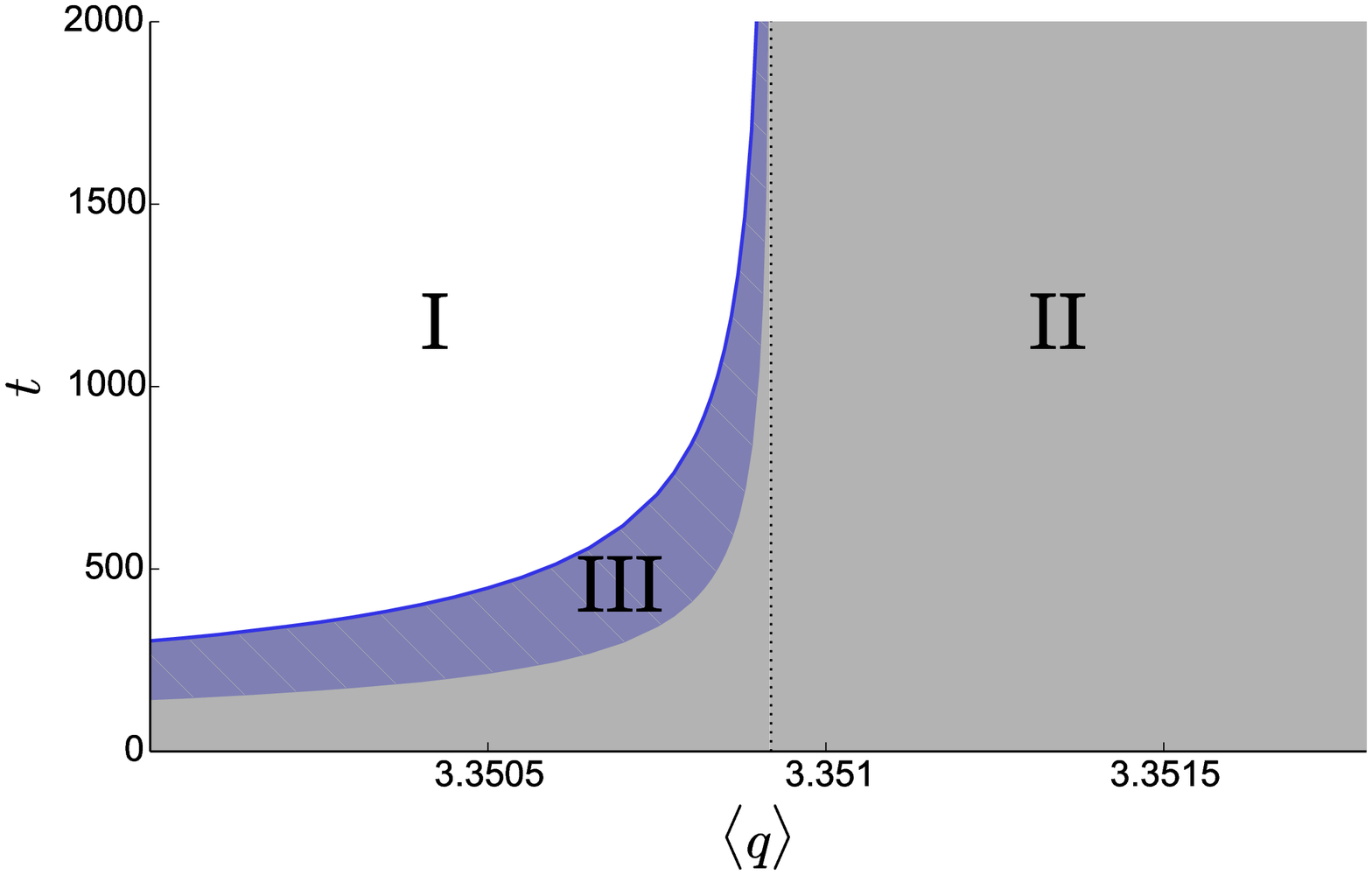}
\caption{(Color online) Phase diagram for the $k$-core pruning process in $\langle q \rangle -t$ plane. The vertical line represent the critical point $\langle q\rangle = \langle q\rangle_c$ ($k=3$ for this figure).
In region~II at $\langle q\rangle > \langle q\rangle_c$, the pruning process reduces the network to the giant $k$-core as time approaches infinity.
Only finite corona clusters are present in region~II.
A giant corona cluster
is present in region~III.
The mean branching is $1$ on the border between regions~II and III.
The mean branching is below $1$ in region II and larger than $1$ in
region III.
The network collapses at times on the upper boundary of region~III so there is no network in region~I.
\label{fig:corona}}
\end{figure}


\section{Branching processes of pruning}
\label{branching processes}

In this section, to understand the nature of the critical dynamics of the pruning process,
we study the spreading of damage through the network and the structural changes during this process.
The probability $\mathcal{P}(n,t)$ of the branching process is given by Eq.~(\ref{branching_dist_exact}) that takes a simple form within the non-crossing approximation,
\begin{equation}
\label{d1}
\mathcal{P}(n,t) = \binom{k-1}{n} s_{t}^{n}(1-s_{t})^{k-1-n}
.
\end{equation}
Since the parameter $s_{t}$ given by Eq.~(\ref{s}) is the probability to encounter a vertex of degree $k$ (corona vertex), the probability Eq. (\ref{d1}) is precisely the probability that following an edge we arrive at a corona vertex which has $n$ corona vertices at the ends of emanating edges \cite{gdm2006}.
It is important to note that as the network evolves according
Eqs.~(\ref{r})--(\ref{P0update}) during the $k$-core pruning process,
so do the corona clusters and hence their size distribution.
Since the
probability $s_t$ and, therefore, the probability $\mathcal{P}(n,t)$
of the branching process depend on time,
the size distribution of branches of removed vertices is therefore
related but not equal to the instantaneous size distribution of corona
clusters (see the following subsection).


\subsection{Branching processes at $\langle q \rangle < \langle q \rangle_c$}
\label{brach-below}

The numerical solution of Eqs.~(\ref{r})--(\ref{P0update}) and simulation show that, in the case $\langle q \rangle < \langle q \rangle_c$ during the plateau stage, the
pruning process
develops as a branching process, as described in
Sec.~\ref{non-crossing appr}.
The branching process of removals evolves in agreement with Eqs.~(\ref{s})--(\ref{k-1-generation}).
Examples of typical
pruning trees
are illustrated in Fig.~\ref{fig:branching_trees} for $k=3$.
The figure
shows that crossings between the branching trees are only abundant at the beginning of the pruning process and are rare in the plateau stage.
The crossings also are abundant at the end of the plateau stage, signaling a collapse in which the entire network disappears.

The full branching distribution, given by
Eq.~(\ref{branching_dist_exact}) is shown in
Fig.~\ref{fig:branching_dist}. It is
similar both above and below
the transition, and barely changes throughout most of the pruning
process. In the figure we also show the branching distribution
observed in simulations. The agreement with theory is good,
however there are noticeable finite size fluctuations, which are
largest when the pruning process is slowest:
this occurs in the middle of the plateau period.
In contrast to this behavior, fluctuations in the case
$\langle q \rangle \geq \langle q\rangle_c$ are enhanced with increasing
time (see Fig. 7). Critical behavior of fluctuations is a common
property of systems approaching the critical point of a continuous
phase transition, or the limiting point of the metastable states of a
first-order phase transition, however, discussion of these phenomena
is beyond the scope of the present paper.

In the early part of the plateau, $P(k-1,t)$ decreases, reaching a
minimum in the middle of the plateau stage, at $t=t_m$.
From Eq. (\ref{B2}) this
corresponds to the point when $b_t$ reaches $1$.
A Taylor expansion around this point (see Appendix \ref{plateau-appr})
gives the temporal behavior of $P(k-1,t)$ in the plateau stage
\begin{equation}
\label{degree distr below}
P(k{-}1,t)= P(k-1,t_m) \Bigl[1+ \frac{1}{2}\Bigl(\frac{t}{t_m} - 1 \Bigr)^2 C_p + \dots \Bigr],
\end{equation}
The corresponding equation for $b_t$ is
\begin{equation}
\label{branching below}
b_t\approx 1+\Bigl(1 - \frac{t}{t_m}\Bigr) C_b,
\end{equation}
Our analysis of the plateau stage in Appendix~\ref{plateau-appr}
shows that $C_b \sim 1/T \propto \sqrt{\langle q \rangle_c - \langle q
  \rangle} \ll 1$. This analytical result agrees with our observation
from the numerical solutions.
The mean branching
$b_t$, Eq.~(\ref{mean_branching exact}), is slightly below $1$ in the
beginning of the plateau stage. As time increases, $b_t$ tends to
increase, as pruning of vertices decreases the mean degree of the network.
The mean branching reaches $1$ at $t_m$ as required, then
continues to increase, with an accelerating rate of pruning, until the network finally collapses rapidly, as seen on the left in Fig.~\ref{fig:mean_branching}.
We observe from numerical solution of the exact
Eqs.~(\ref{r})--(\ref{P0update}), and from simulations, that the
minimum occurs in the middle
of the plateau stage, i.e. $t_m = T/2$, see Fig. \ref{fig:evolution}.
The numerical solution of exact Eqs.~(\ref{r})--(\ref{P0update}) shows that the coefficient $C_p$ is of order 1. Using this result and Eq.~(\ref{C5}) in Appendix~\ref{plateau-appr}, we find a relationship between $P(k-1, t_m)$ and the plateau duration $T$,
\begin{equation}
P(k-1, t_m) \sim \frac{1}{T^2} \propto \langle q \rangle_c - \langle q \rangle
.
\label{minP(k-1)}
\end{equation}

The instantaneous size distribution $\Pi(S,t)$ of finite corona clusters can be found directly from the degree distribution $P(q,t)$ at every time $t$:
\begin{equation}
\label{size distribution}
\Pi(S,t)=CS^{-3/2}e^{-S/S^*(t)}
\end{equation}
where  $S^*(t) \rightarrow \infty$ at the critical point of the emergence of a giant corona cluster.
According to \cite{gdm2006}, a giant connected cluster of corona vertices is present when
\begin{equation}
\label{giant_corona}
b_t=\frac{ (k-1)k P(k,t)}{ \langle q\rangle_t} \geq 1
.
\end{equation}
In the case of $k=3$, we have $S^*(t) =-1/\ln[4s_t(1-s_t)]$ where
$s_t=3P(3,t)/\langle q \rangle_t$ according to Eq. (\ref{s})
\cite{gdm2006}. In Appendix~\ref{plateau-appr} we show that a giant
corona cluster appears continuously at the same time $t_m$ when the fraction $P(k-1,t)$ of $k-1$ nodes achieves a minimum.  Such a giant corona cluster will be
consumed by the pruning process, guaranteeing the collapse of the whole
network in finite time. A similar continuous emergence of a giant subgraph prone to failure was recently observed in interdependent networks in Ref.~\cite{zhou2014simultaneous}.
The left side of Eq.~(\ref{giant_corona}) is identical to
Eq.~(\ref{mean_branching}), so the border of the region where a
giant corona cluster appears is at the point where the mean branching
of the pruning process equals 1. The region in the $\langle q\rangle -t$ plane where the giant corona cluster is present is marked in Fig.~\ref{fig:corona} as region
$\mbox{III}$.
Note that a giant corona cluster only appears below $\langle
q\rangle_c$ in the plateau stage.
At $\langle q\rangle=\langle q\rangle_c$,
at any time $t$ there are only finite corona clusters. When $t \to
\infty$,  the size distribution of corona clusters tends to the power
law function Eq. (24), corresponding to the critical point of the
emergence of a giant corona cluster.
Above $\langle q\rangle_c$, there are only finite corona clusters at
any time.


\subsection{Branching processes at $\langle q \rangle > \langle q \rangle_c$}
\label{braching-above}

Above the transition point, with increasing time the degree distribution $P(q,t)$ tends to the steady distribution $P_{k}(q)$  with mean degree $\langle q \rangle_k =\sum_{q\geq k }q P_{k}(q)$ while $P(k-1,t) \rightarrow 0$.
In turn, the mean branching $b_t$ saturates at a
constant value $b_{k}$ less than 1 (see the right side of
Fig.~\ref{fig:mean_branching}).
If $\langle q \rangle$ is close to $\langle q \rangle_c$,
$1 - k(k-1)P_{k}(k)/\langle q \rangle_k \approx B\sqrt{\langle q
  \rangle_c - \langle q \rangle}$ where $B$ is a constant
\cite{gdm2006}, and using Eq.~(\ref{mean_branching}) we have $b_{k}
\approx 1 -B\sqrt{\langle q \rangle_c - \langle q \rangle}$.
Substituting the constant $b_k$ for $b_t = k(k-1)P_{k}(k,t)/\langle q
\rangle_t$ in
 Eq.~(\ref{B2}) in Appendix~\ref{crit-relax-appr}, and solving, we find an
 exponential decay of $P(k-1,t)$, Eq.~(\ref{relaxation above}),  and a
 relationship between the relaxation time $\tau$ and the branching
 coefficient $b_{k}$,
\begin{equation}
\label{tau_b_t}
b_{k} =1 - \tau^{-1}
.
\end{equation}
Therefore,
\begin{equation}
\label{tau_above}
\tau^{-1}= 1 - k(k-1)P_{k}(k)/\langle q \rangle_k \approx B\sqrt{\langle q \rangle_c - \langle q \rangle }
\end{equation}
in agreement with the numerical solution Eq.~(\ref{tau_divergence}).
The pruning process
only evolves within finite corona clusters, and the network survives at any time $t$ (the region $\mbox{II}$ in Fig.~\ref{fig:corona}) approaching the steady $k$-core as time approaches infinity.


\subsection{Critical branching process}

Exactly at the critical point, $\langle q \rangle = \langle q
\rangle_c$, the branching $b_t$ comes arbitrarily close to 1,
but only reaches that value in the infinite time limit.
The leading term in $1-b_t$ is a monotonically decreasing function of
$t$. Solving
Eqs.~(\ref{B1})--(\ref{B2}) in Appendix~\ref{crit-relax-appr}, we find that the function $P(k-1,t)$ has power-law behavior, Eq.~(\ref{critpowerlaw}), with critical exponent $\sigma=2$. This behavior corresponds to the mean branching $b_t$ increasing as
\begin{equation}
\label{critical branching}
b_t = 1-2/t +O(1/t^2)
.
\end{equation}
This kind of time dependence of the mean branching is known to lead to the avalanche lifetime distribution  $L(\mathcal{T})\propto \mathcal{T}^{-2}$ \cite{Harris1989} found in various models (see, for example, \cite{zlbs1995,sethna2001crackling}) and real systems (for example, in the brain \cite{beggs2003neuronal}).
This suggests  that the power-law relaxation Eq.~(\ref{critpowerlaw})
and the avalanche lifetime distribution
have the same origin.

Since the mean branching $b_t=(k-1) s_t$ tends to 1 when $t\rightarrow \infty$, we have $s_t \rightarrow 1/(k-1)$. Equation (\ref{d1}) gives the following exact result:
\begin{equation}
\label{d4}
\mathcal{P}(n,\infty) = \binom{k-1}{n} \frac{(k-2)^{k-1-n}}{(k-1)^{k-1}}
.
\end{equation}
In the case of $k=3$, we obtain
$\mathcal{P}(0,\infty)=\mathcal{P}(2,\infty)= 1/4$ and
$\mathcal{P}(1,\infty)=1/2$. These values agree with results obtained
by our simulations and numerical solutions that are displayed in
Fig.~\ref{fig:branching_dist}.


\section{Discussions and Conclusions}

In this paper we
have developed the theory of the
$k$-core pruning process in uncorrelated, sparse random networks.
Employing the numerical solution of
the exact
evolution equations, Eqs.~(\ref{r})--(\ref{P0update}), an asymptotic
analysis, and simulations
in \er graphs, we revealed that this process demonstrates three
different kinds of critical behavior depending on whether the mean
degree $\langle q \rangle$ of the initial network is above, equal to, or below the critical point, $\langle q \rangle_c$, corresponding to the emergence of the giant $k$-core.
We found that above the critical point, at large times
the network relaxes exponentially to the steady $k$-core.
At the critical point, $\langle q \rangle= \langle q \rangle_c$, the dynamics is critical and it is described by a power law with respect to time ($\propto 1/t^2$).
Below the critical point, the pruning eliminates an infinite network in a finite time.
The duration of the transient process diverges when $\langle q \rangle$ tends to $\langle q \rangle_c$ from below.

We found mechanisms for these critical phenomena.
Studying the structure of paths along which damage is spreading in the
network, we found that the damage spreading is a branching process.
Our analysis showed that it is the evolving clusters of
nodes of degree $k$ (``corona clusters'') that provide the substrate
for the branching process. Indeed, if a vertex of degree $k$ loses an
edge and is removed, then this removal triggers a removal of all
corona vertices, one by one, which belong to the same corona cluster.
Using analytical methods and simulation,
we showed that the pruning
can be considered as a branching process that begins after a short
initial period of rapid network change.
During this process, independent branching trees develop with
branching ratio close to $1$.
The temporal behavior of the mean branching plays a crucial
role in the branching process and slowdown of the $k$-core pruning dynamics
at the critical point and during the plateau stage.
To understand the branching process it is important to note
that corona clusters evolve in time. When damage propagates over the
network, on one hand, it removes corona nodes, but on the
other hand, it decreases the degrees of neighboring nodes, producing new
corona vertices and thus increasing  size of other corona clusters,
which can be pruned at a later time. Due to this, the branching
probability becomes time dependent.
The mean branching is close to $1$
during the whole plateau stage, below $\langle q \rangle_c$.
At the beginning of the stage the mean branching is a little bit
smaller than $1$, and slowly increases with time.
It
reaches the value $1$ approximately at the middle of this stage and then continues to increase.
 At this point a giant corona cluster is formed providing a substrate for the complete collapse of the network at the end of the plateau stage.
Exactly at the critical point, the mean branching comes arbitrarily
close to $1$, but never reaches it at any finite time. This leads to a
powerlaw decay in the fraction of nodes of degree $k-1$. The branching
trees of pruning become arbitrarily long, but a giant corona cluster
is never formed, until $t \to \infty$.
Finally, we found that above the critical point $\langle q \rangle_c$,
the mean branching saturates at a constant value less than 1. In this
case, mean size of branches is finite and relaxation to the steady
$k$-core follows an exponential law.

The $k$-core pruning process in sparse, uncorrelated random complex networks is a representative model of dynamics in complex systems undergoing hybrid phase transitions.
We have solved this model and have developed the
complete description of critical dynamical phenomena including the long-lasting transient process, critical relaxation, and critical slowing down. We suggest that our results could be useful for understanding similar collective phenomena that occur in other complex systems near
discontinuous (hybrid and first-order) phase transitions.


\section{Acknowledgements}
This work is funded by FEDER funds through the COMPETE 2020 Programme and National Funds through FCT - Portuguese Foundation for Science and Technology under the project UID/CTM/50025/2013. This work was partially supported by the FET proactive IP project MULTIPLEX 317532, the FCT project EXPL/FIS-NAN/1275/2013, and the project ``New Strategies Applied to Neuropathological Disorders'' (CENTRO-07-ST24-FEDER-002034) cofunded by QREN and EU. KEL was supported by the FCT Grant No. SFRH/ BPD/ 71883/2010. GJB was supported by the FCT grant No. SFRH/BPD/74040/2010.


\appendix


\section{Relaxation in 1D system near border of metastability}
\label{appendix}

\begin{figure}[t]
\includegraphics[width=0.32\columnwidth]{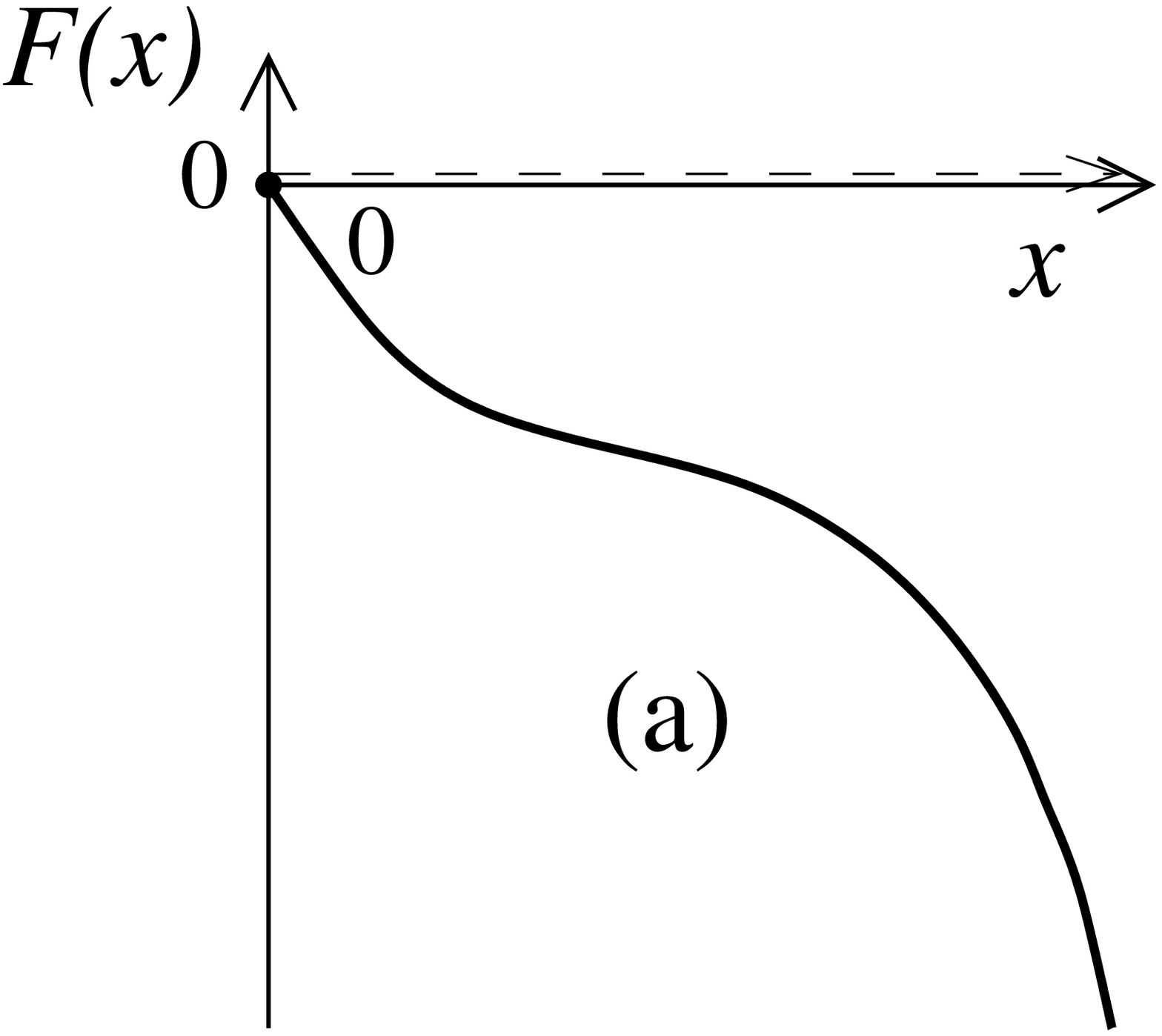}
\includegraphics[width=0.32\columnwidth]{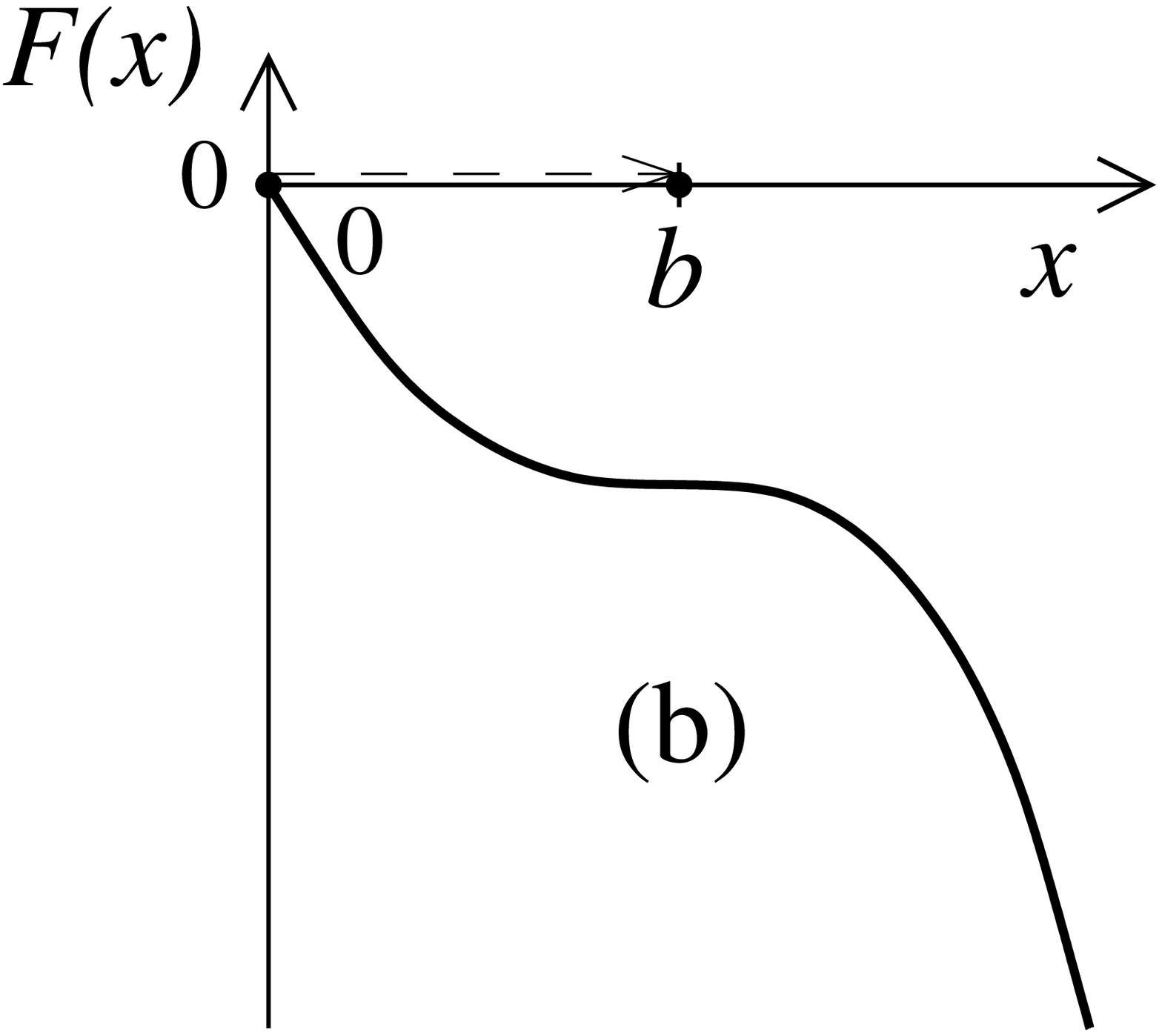}
\includegraphics[width=0.32\columnwidth]{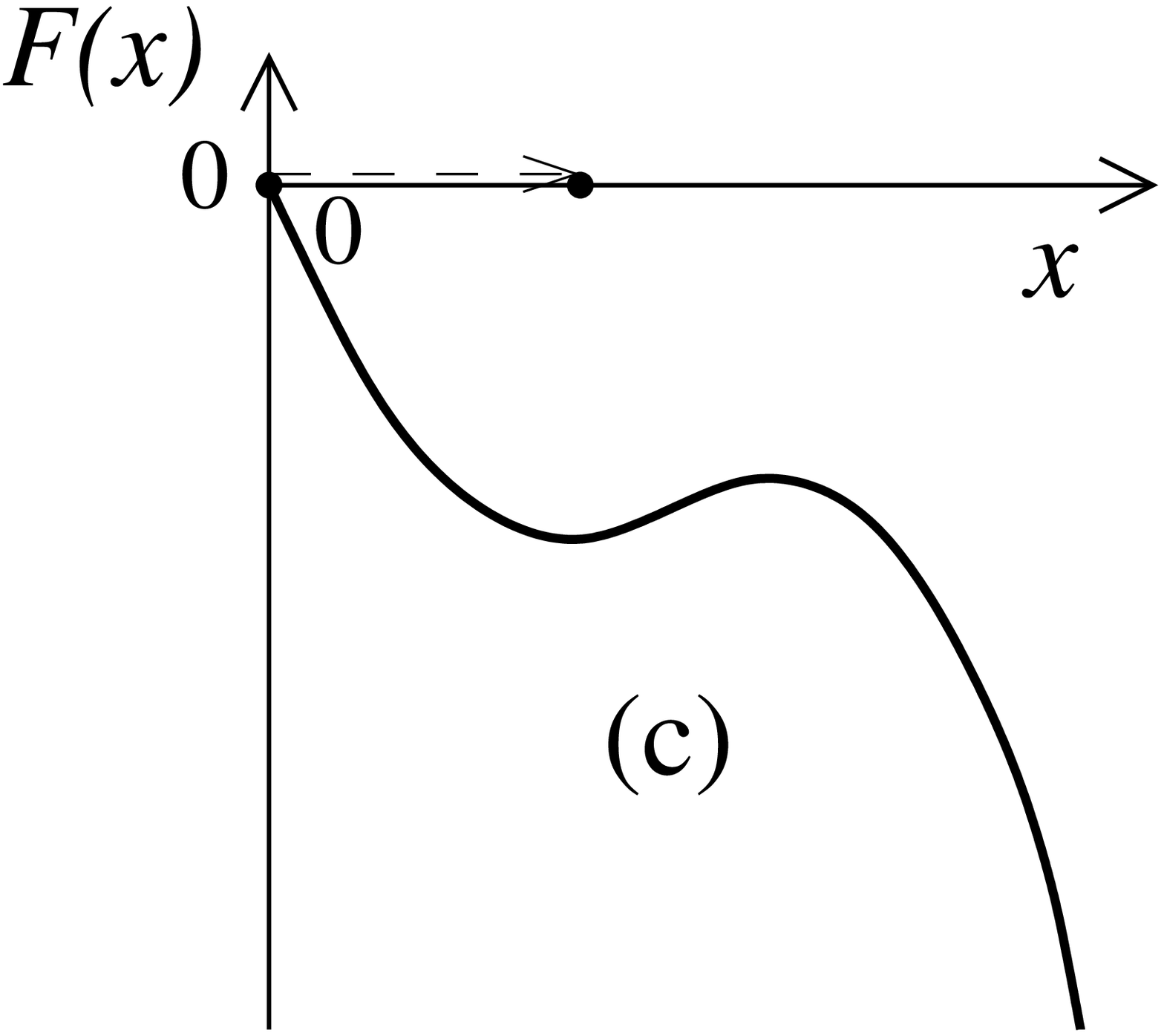}
\caption{
Potential $F(x)$ versus $x$ in Eq.~(\ref{a1}) in the following cases: a) the coefficient $a>a_c \equiv b^2$,  $F'(x)<0$ at any $x$.
b) $a=a_c$, $F'(x=b)=0$ and there is a saddle at $x=b$.
c) $a<a_c$, $F(x)$ has a local minimum and a local maximum.
}
\label{fa1}
\end{figure}

The behavior described in this paper for the $k$-core pruning process
is common for dynamical systems having a saddle point for some set of
system parameters.
Here we consider the simplest dynamical model of this sort, namely a
particle moving in a one-dimensional  potential $F(x)$ in a viscous
medium,  demonstrating
features similar to the $k$-core pruning
process:
\begin{eqnarray}
&&
\partial_t x = - \partial_x F(x)
,
\nonumber
\\[5pt]
&&
F(x) = - a x + b x^2 - \frac{1}{3}x^3
.
\label{a1}
\end{eqnarray}
Here the coefficients $a$ and $b$ are positive, and the variable
(particle's coordinate) $x(t) \geq 0$. The initial condition is
$x(t=0)=0$.
There are three distinct regimes, see Fig.~\ref{fa1}:

\begin{itemize}

\item[(a)]
$a>a_c=b^2$, normal phase, with $\partial_x F(x)<0$ at any $x$. At the
  end of the process $x$ approaches infinity;

\item[(b)]
$a=a_c$, resulting in the saddle point $x_s=b$ in $F(x)$. As $t\to\infty$, $x(t)$ approaches $b$;

\item[(c)]
$a<a_c=b^2$, which gives the local minimum (``metastable state'') at
  $x_m = b - \sqrt{b^2-a}$. As $t\to\infty$, $x(t)$ approaches $x_m$.

\end{itemize}

The straightforward solution of Eq.~(\ref{a1}) shows that in the
normal phase [regime (a)], i.e., when $a>a_c$, the variable $x$,
starting from $0$, approaches infinity in a finite time $T(a)$. For
$a$ close to $a_c$, the process greatly slows down when $x(t)$ passes
the value $b$, and we obtain the asymptotic expression
\begin{equation}
T \cong 2 \int_0^b \frac{dx}{a-2bx+x^2} \cong \frac{\pi}{\sqrt{a-b^2}}
.
\label{a2}
\end{equation}
This time diverges at the critical point $a_c=b^2$ [regime (b)], at
which $x$ relaxes to the saddle point value $b$ slowly, in a power-law
way. Asymptotically, we get
\begin{equation}
x(t)-b
\cong - \frac{1}{b^3}\,\frac{1}{t}
.
\label{a3}
\end{equation}
Note that in the case $F(x)=c x^4$, which corresponds to the second
order phase transition within the Landau
theory,  the
equation $\partial_t x = - \partial_x F(x)$
leads to the critical relaxation $x \propto t^{-1/2}$.

In regime (c), $x(t)$ relaxes exponentially to the local minimum value $x_m=b - \sqrt{b^2-a}$. Asymptotically,
\begin{equation}
x(t)-
x_m
\propto 
e^{-t/\tau}
.
\label{a4}
\end{equation}
Here $\tau$ is the relaxation time,
\begin{equation}
\tau = \frac{1}{2\sqrt{b^2-a}}
\label{a5}
\end{equation}
diverging at the critical point.

The square root critical singularities of $T$ and $\tau$,
Eqs.~(\ref{a2}) and (\ref{a5}), respectively, coincide with those of
the $k$-core pruning process, Eqs.~(\ref{T_divergence}) and (\ref{tau_divergence}).
From these expressions
we obtain the remarkably beautiful ratio of the critical amplitudes of $T$ and $\tau$:
\begin{equation}
\frac{T(a{-}a_c)}{\tau(a_c{-}a)} = 2\pi
\label{a6}
\end{equation}
coinciding with the corresponding ratio that we found for the $k$-core pruning.

Near the critical point, the time $T$ to complete the process (to run away from $x=0$ to infinity) in the normal phase is strongly influenced by small variations of the parameters of the system.
To quantify this effect, we introduce a time dependent perturbation $h(t)$ of the coefficient $a$ in the potential $F(x)$, namely, $a-h(t)$.
Let $h(t)$ be a constant $h$ within the interval of width $\epsilon$ around some moment $\tilde{t}$.
We define the response of $T$ to $h$ as
\begin{equation}
\chi(\tilde{t})  \equiv \lim_{h,\epsilon\to 0} \frac{\Delta T(h,\epsilon,\tilde{t})}{h\epsilon}
,
\label{a7}
\end{equation}
where $\Delta T(h,\epsilon,\tilde{t})=T(h,\epsilon,\tilde{t})-T(h{=}0,\epsilon{=}0,\tilde{t})$ is the variation of the time $T$
 due to the perturbation $h(t)$.
This response takes an elegant asymptotic form
as $a$ approaches the critical point $a_c=b^2$,
\begin{equation}
\chi(\tilde{t}) \cong \frac{1}{a-b^2}\Bigg[\Big(\frac{\pi}{2}\Big)^2 \Big(\frac{\tilde{t}}{T/2}-1\Big)^2 + 1\Bigg]^{-1}
,
\label{a8}
\end{equation}
which has a Lorentz shape in terms of the moment of the perturbation, $\tilde{t}$, and diverges according to the Curie-Weiss law.
Here $T$ is the time, given by Eq. (\ref{a2}), to run away to infinity in the absence of
perturbation, $h=0$, and $T/2$ is the time at which the particle
passes the point $x=b$. This divergence of the response $\chi$ at the
critical point indicates the presence of strong fluctuations near the
critical point, which we observe in the $k$-core pruning process (see
Figs. \ref{fig:branching_dist} and \ref{fig:mean_branching}).


\section{Critical relaxation in the non-crossing approximation}
\label{crit-relax-appr}

Let us solve Eqs.~(\ref{q-generation})--(\ref{mean-degree-appr}) and find the critical behavior of $P(k-1,t)$ at the critical point $\langle q \rangle =\langle q \rangle_c$. At $t\gg 1$, we consider $P(q,t)$ as a function of continuous time $t$. In this limit,
Eqs.~(\ref{q-generation})--(\ref{mean-degree-appr}) take a differential form,
\begin{eqnarray}
&&
\!\!\!\!\!\!\!\!\!\!\!\!\!\!\!\!\!\!\!\!\!\!\!\!\!\!\!\!\!\!\!\!\!\!\!\!\!\!\!
\frac{\partial P(q,t)}{\partial t} {=} \frac{ k{-}1}{ \langle q\rangle_t}  P(k{-}1,t)\! \Bigl[(q{+}1)P(q{+}1,t) {-} q P(q,t) \Bigr]
,
\label{B1}
\\[5pt]
\frac{\partial P(k{-}1,t)}{\partial t}&{=}& \left[\frac{ k(k{-}1)}{
    \langle q\rangle_t} P(k,t) -1  \right]  P(k{-}1,t),~~~~
\label{B2}
\\[5pt]
\frac{\partial P(0,t)}{\partial t}&{=}& P(k-1,t),
\label{B2b}
\\[5pt]
\langle q\rangle_t &{=}& (k-1)P(k-1,t) + \sum_{q\geq k} q P(q,t)
,
\label{B3}
\end{eqnarray}
where $q \geq k$. In order to solve these equations, we use the fact that with increasing $t$ the degree distribution $P(q,t)$ for $q\geq k$ tends to the steady
degree distribution of the $k$-core, $P_{k}(q)$, i.e., $P(q,t)=P_{k}(q) + \delta P(q,t)$. Moreover, $\delta P(q,t) \rightarrow 0$ and  $P(k-1,t) \rightarrow 0$ in the infinite time limit. At the critical point, the distribution satisfies the condition $k(k-1)P_{k}(k)=\langle q \rangle_k$ where  $\langle q \rangle_k \equiv \sum_{q\geq k }q P_{k}(q)$ \cite{dgm2006,gdm2006}.
We solve Eq.~(\ref{B1}) in the first order in $P(k-1,t)$ and find
\begin{equation}
\delta P(q,t){=}\frac{(k{-}1)}{\langle q \rangle_k}\Bigl[ q P_{k}(q) {-}(q{+}1)P_{k}(q{+}1) \Bigr] \!\! \int_{t}^{\infty} \!\!\!\!\!\!\! P(k{-}1,t) dt
.
\label{B4}
\end{equation}
Then, using Eq.~(\ref{B3}), we find $\langle q\rangle_t$ in the first order in $P(k-1,t)$. Substituting these results into Eq.~(\ref{B2}) gives an equation,
\begin{eqnarray}
\frac{\partial P(k{-}1,t)}{\partial t}&=& - vP(k-1,t)\int_{t}^{\infty} P(k-1,t) dt
\nonumber
\\[5pt]
&-&\!\frac{(k{-}1)}{\langle q \rangle_{k}} P^{2}(k{-}1,t) {+} O(P^{3}(k{-}1,t))
,~~~~~
\label{B5}
\end{eqnarray}
where
\begin{equation}
\! v= \frac{k(k{-}1)^2 (k{+}1)}{\langle q \rangle_{k}^{2}}P_{k}(k{+}1){-}\frac{(k{-}1)(k-2)}{\langle q \rangle_{k}}
.
\label{B6}
\end{equation}
Equation (\ref{B4}) has a solution
\begin{equation}
P(k-1,t) = \frac{2}{vt^2} + O(1/t^3)
.
\label{B7}
\end{equation}
Numerical solution of the exact Eqs.~(\ref{r})--(\ref{P0update}) confirms this result (see Fig.~\ref{fig:powerlaw}).
Using Eqs.~(\ref{B4}) and  (\ref{B7}), we find the mean branching,
\begin{equation}
b_t = \frac{k(k-1)P(k,t)}{\langle q \rangle_t} \approx 1 - \frac{2}{t} + O(1/t^2)
.
\label{B8}
\end{equation}
Note that this is the universal critical behavior of branching processes \cite{Harris1989}.


\section{Plateau stage in the non-crossing approximation}
\label{plateau-appr}

If $\langle q \rangle {<}\langle q \rangle_c$, with increasing time the fraction $P(k{-}1,t)$ of nodes of degree $k-1$ achieves a minimum at a time $t_m$ in the middle of the plateau stage (see Fig.~\ref{fig:evolution}). The time $t_m$ is determined by the condition
\begin{equation}
\frac{\partial P(k-1,t)}{\partial t} \Big|_{t=t_m} = 0
.
\label{C1}
\end{equation}
According to Eq.~(\ref{B2}), at $t=t_m$ the following equality also holds,
\begin{equation}
\frac{k(k-1)P(k,t_m)}{\langle q \rangle_{t_m}} = 1
.
\label{C2}
\end{equation}
It signals the percolation of corona clusters [see Eq.~(\ref{giant_corona})]. Thus, the minimum of  $P(k-1,t)$ occurs
when the giant corona cluster appears. Near the minimum, we can use the Taylor expansion
\begin{equation}
P(k{-}1,t)= P(k-1,t_m) \Bigl[1+ \frac{1}{2}\Bigl(\frac{t}{t_m} - 1 \Bigr)^2 C_p + \dots \Bigr]
,
\label{C3}
\end{equation}
where
\begin{equation}
C_p \equiv \frac{t^{2}_m}{P(k-1,t_m)} \frac{\partial ^2 P(k{-}1,t)}{\partial t^2}\Big|_{t=t_m}
.
\label{C4}
\end{equation}
Differentiating Eq.~(\ref{B2}) with respect to $t$, we find the second derivative and
\begin{equation}
C_p = t^{2}_m v_m P(k-1,t_m)
,
\label{C5}
\end{equation}
where
\begin{equation}
v_m = \frac{k(k-1)^2 (k+1)}{\langle q \rangle_{t_m}^{2}}P(k+1, t_m) - \frac{(k-1)(k-2)}{\langle q \rangle_{t_m}}
.
\label{C6}
\end{equation}
We estimate $P(k-1, t_m)$ and $t_m$ using the numerical solution of exact Eqs.~(\ref{r})--(\ref{P0update}). Our numerical results in Secs.~\ref{plateau-stage} and \ref{brach-below}  show that the coefficient $C_p$ is of the order of $1$,
and $t_m \approx T/2$, i.e., the minimum takes place at the middle of the plateau stage. Equation (\ref{C5}) gives a relationship between $P(k-1, t_m)$ and
the duration $T$ of the entire pruning process,
\begin{equation}
P(k-1, t_m) \sim \frac{1}{T^2} \propto \langle q \rangle_c - \langle q \rangle
.
\label{C7}
\end{equation}
Note that in the neighborhood of the threshold $\langle q \rangle_c$,
the plateau duration tends to the time $T$ to complete the pruning process.
Equation~(\ref{mean_branching}) and the Taylor expansion of the
function $P(k,t)$ give the temporal behavior of the mean branching $b_t$ near $t_m$,
\begin{equation}
b_t= 1 + \Bigl(\frac{t}{t_m} - 1 \Bigr) C_b + \dots
,
\label{C8}
\end{equation}
where
\begin{eqnarray}
C_b &=& t_m \frac{k(k-1)}{\langle q \rangle_{t_m}^2}P(k-1,t_m)
\nonumber
\\[5pt]
&\times& \Bigl[(k{-}1)^2 P(k{+}1,t_m) {+} 2(k{-}1){-}\langle q \rangle_{t_m} \Bigl]
.
\label{C9}
\end{eqnarray}
This equation shows that $C_b$ is small since $C_b \sim 1/T \propto
\sqrt{\langle q \rangle_c - \langle q \rangle} \ll 1$. This result is
also supported by our numerical solution and simulations for
Erd\H{o}s--R\'enyi graphs (see Sec.~\ref{brach-below}).


\bibliography{bibliography}

\begin{thebibliography}{29}%
\makeatletter
\providecommand \@ifxundefined [1]{%
 \@ifx{#1\undefined}
}%
\providecommand \@ifnum [1]{%
 \ifnum #1\expandafter \@firstoftwo
 \else \expandafter \@secondoftwo
 \fi
}%
\providecommand \@ifx [1]{%
 \ifx #1\expandafter \@firstoftwo
 \else \expandafter \@secondoftwo
 \fi
}%
\providecommand \natexlab [1]{#1}%
\providecommand \enquote  [1]{``#1''}%
\providecommand \bibnamefont  [1]{#1}%
\providecommand \bibfnamefont [1]{#1}%
\providecommand \citenamefont [1]{#1}%
\providecommand \href@noop [0]{\@secondoftwo}%
\providecommand \href [0]{\begingroup \@sanitize@url \@href}%
\providecommand \@href[1]{\@@startlink{#1}\@@href}%
\providecommand \@@href[1]{\endgroup#1\@@endlink}%
\providecommand \@sanitize@url [0]{\catcode `\\12\catcode `\$12\catcode
  `\&12\catcode `\#12\catcode `\^12\catcode `\_12\catcode `\%12\relax}%
\providecommand \@@startlink[1]{}%
\providecommand \@@endlink[0]{}%
\providecommand \url  [0]{\begingroup\@sanitize@url \@url }%
\providecommand \@url [1]{\endgroup\@href {#1}{\urlprefix }}%
\providecommand \urlprefix  [0]{URL }%
\providecommand \Eprint [0]{\href }%
\providecommand \doibase [0]{http://dx.doi.org/}%
\providecommand \selectlanguage [0]{\@gobble}%
\providecommand \bibinfo  [0]{\@secondoftwo}%
\providecommand \bibfield  [0]{\@secondoftwo}%
\providecommand \translation [1]{[#1]}%
\providecommand \BibitemOpen [0]{}%
\providecommand \bibitemStop [0]{}%
\providecommand \bibitemNoStop [0]{.\EOS\space}%
\providecommand \EOS [0]{\spacefactor3000\relax}%
\providecommand \BibitemShut  [1]{\csname bibitem#1\endcsname}%
\let\auto@bib@innerbib\@empty
\bibitem [{\citenamefont {Seidman}(1983)}]{seidman1983network}%
  \BibitemOpen
  \bibfield  {author} {\bibinfo {author} {\bibfnamefont {S.~B.}\ \bibnamefont
  {Seidman}},\ }\href@noop {} {\bibfield  {journal} {\bibinfo  {journal}
  {Social Networks}\ }\textbf {\bibinfo {volume} {5}},\ \bibinfo {pages} {269}
  (\bibinfo {year} {1983})}\BibitemShut {NoStop}%
\bibitem [{\citenamefont {Bollob{\'a}s}(1984)}]{bollobas1984}%
  \BibitemOpen
  \bibfield  {author} {\bibinfo {author} {\bibfnamefont {B.}~\bibnamefont
  {Bollob{\'a}s}},\ }in\ \href@noop {} {\emph {\bibinfo {booktitle} {Graph
  Theory and Combinatorics (Cambridge 1983)}}}\ (\bibinfo  {publisher}
  {Academic Press},\ \bibinfo {address} {London},\ \bibinfo {year} {1984})\
  pp.\ \bibinfo {pages} {35--57}\BibitemShut {NoStop}%
\bibitem [{\citenamefont {{\L}uczak}(1991)}]{luczak1991size}%
  \BibitemOpen
  \bibfield  {author} {\bibinfo {author} {\bibfnamefont {T.}~\bibnamefont
  {{\L}uczak}},\ }\href@noop {} {\bibfield  {journal} {\bibinfo  {journal}
  {Discrete Mathematics}\ }\textbf {\bibinfo {volume} {91}},\ \bibinfo {pages}
  {61} (\bibinfo {year} {1991})}\BibitemShut {NoStop}%
\bibitem [{\citenamefont {Batagelj}\ and\ \citenamefont
  {Zaver{\v{s}}nik}(2002)}]{batagelj2002generalized}%
  \BibitemOpen
  \bibfield  {author} {\bibinfo {author} {\bibfnamefont {V.}~\bibnamefont
  {Batagelj}}\ and\ \bibinfo {author} {\bibfnamefont {M.}~\bibnamefont
  {Zaver{\v{s}}nik}},\ }\href@noop {} {\bibfield  {journal} {\bibinfo
  {journal} {arXiv:cs/0202039}\ } (\bibinfo {year} {2002})}\BibitemShut
  {NoStop}%
\bibitem [{\citenamefont {Dorogovtsev}\ \emph
  {et~al.}(2006{\natexlab{a}})\citenamefont {Dorogovtsev}, \citenamefont
  {Goltsev},\ and\ \citenamefont {Mendes}}]{dgm2006}%
  \BibitemOpen
  \bibfield  {author} {\bibinfo {author} {\bibfnamefont {S.~N.}\ \bibnamefont
  {Dorogovtsev}}, \bibinfo {author} {\bibfnamefont {A.~V.}\ \bibnamefont
  {Goltsev}}, \ and\ \bibinfo {author} {\bibfnamefont {J.~F.~F.}\ \bibnamefont
  {Mendes}},\ }\href {\doibase 10.1103/PhysRevLett.96.040601} {\bibfield
  {journal} {\bibinfo  {journal} {Phys. Rev. Lett.}\ }\textbf {\bibinfo
  {volume} {96}},\ \bibinfo {pages} {040601} (\bibinfo {year}
  {2006}{\natexlab{a}})}\BibitemShut {NoStop}%
\bibitem [{\citenamefont {Buldyrev}\ \emph {et~al.}(2010)\citenamefont
  {Buldyrev}, \citenamefont {Parshani}, \citenamefont {Paul}, \citenamefont
  {Stanley},\ and\ \citenamefont {Havlin}}]{Buldyrev2010}%
  \BibitemOpen
  \bibfield  {author} {\bibinfo {author} {\bibfnamefont {S.}~\bibnamefont
  {Buldyrev}}, \bibinfo {author} {\bibfnamefont {R.}~\bibnamefont {Parshani}},
  \bibinfo {author} {\bibfnamefont {R.}~\bibnamefont {Paul}}, \bibinfo {author}
  {\bibfnamefont {H.~E.}\ \bibnamefont {Stanley}}, \ and\ \bibinfo {author}
  {\bibfnamefont {S.}~\bibnamefont {Havlin}},\ }\href {\doibase
  10.1038/nature08932} {\bibfield  {journal} {\bibinfo  {journal} {Nature}\
  }\textbf {\bibinfo {volume} {464}},\ \bibinfo {pages} {1025} (\bibinfo {year}
  {2010})}\BibitemShut {NoStop}%
\bibitem [{\citenamefont {Albert}\ and\ \citenamefont
  {Barab{\'a}si}(2002)}]{ab2002}%
  \BibitemOpen
  \bibfield  {author} {\bibinfo {author} {\bibfnamefont {R.}~\bibnamefont
  {Albert}}\ and\ \bibinfo {author} {\bibfnamefont {A.-L.}\ \bibnamefont
  {Barab{\'a}si}},\ }\href@noop {} {\bibfield  {journal} {\bibinfo  {journal}
  {Rev. Mod. Phys.}\ }\textbf {\bibinfo {volume} {74}},\ \bibinfo {pages} {47}
  (\bibinfo {year} {2002})}\BibitemShut {NoStop}%
\bibitem [{\citenamefont {Newman}(2003)}]{Newman2003}%
  \BibitemOpen
  \bibfield  {author} {\bibinfo {author} {\bibfnamefont {M.~E.~J.}\
  \bibnamefont {Newman}},\ }\href {\doibase 10.1137/S003614450342480}
  {\bibfield  {journal} {\bibinfo  {journal} {SIAM Review}\ }\textbf {\bibinfo
  {volume} {45}},\ \bibinfo {pages} {167} (\bibinfo {year} {2003})}\BibitemShut
  {NoStop}%
\bibitem [{\citenamefont {Dorogovtsev}\ \emph {et~al.}(2008)\citenamefont
  {Dorogovtsev}, \citenamefont {Goltsev},\ and\ \citenamefont
  {Mendes}}]{dgm2008}%
  \BibitemOpen
  \bibfield  {author} {\bibinfo {author} {\bibfnamefont {S.~N.}\ \bibnamefont
  {Dorogovtsev}}, \bibinfo {author} {\bibfnamefont {A.~V.}\ \bibnamefont
  {Goltsev}}, \ and\ \bibinfo {author} {\bibfnamefont {J.~F.~F.}\ \bibnamefont
  {Mendes}},\ }\href {\doibase 10.1103/RevModPhys.80.1275} {\bibfield
  {journal} {\bibinfo  {journal} {Rev. Mod. Phys.}\ }\textbf {\bibinfo {volume}
  {80}},\ \bibinfo {pages} {1275} (\bibinfo {year} {2008})}\BibitemShut
  {NoStop}%
\bibitem [{\citenamefont {Pittel}\ \emph {et~al.}(1996)\citenamefont {Pittel},
  \citenamefont {Spencer},\ and\ \citenamefont {Wormald}}]{pittel1996}%
  \BibitemOpen
  \bibfield  {author} {\bibinfo {author} {\bibfnamefont {B.}~\bibnamefont
  {Pittel}}, \bibinfo {author} {\bibfnamefont {J.}~\bibnamefont {Spencer}}, \
  and\ \bibinfo {author} {\bibfnamefont {N.}~\bibnamefont {Wormald}},\
  }\href@noop {} {\bibfield  {journal} {\bibinfo  {journal} {Journal of
  Combinatorial Theory, Ser. B}\ }\textbf {\bibinfo {volume} {67}},\ \bibinfo
  {pages} {111} (\bibinfo {year} {1996})}\BibitemShut {NoStop}%
\bibitem [{\citenamefont {Bauer}\ and\ \citenamefont
  {Golinelli}(2001)}]{bauer2001core}%
  \BibitemOpen
  \bibfield  {author} {\bibinfo {author} {\bibfnamefont {M.}~\bibnamefont
  {Bauer}}\ and\ \bibinfo {author} {\bibfnamefont {O.}~\bibnamefont
  {Golinelli}},\ }\href@noop {} {\bibfield  {journal} {\bibinfo  {journal}
  {Eur. Phys. J. B: Cond. Matter and Complex Systems}\ }\textbf {\bibinfo
  {volume} {24}},\ \bibinfo {pages} {339} (\bibinfo {year} {2001})}\BibitemShut
  {NoStop}%
\bibitem [{\citenamefont {Zhou}\ \emph {et~al.}(2012)\citenamefont {Zhou},
  \citenamefont {Bashan}, \citenamefont {Berezin}, \citenamefont {Cohen},\ and\
  \citenamefont {Havlin}}]{zhou2012dynamics}%
  \BibitemOpen
  \bibfield  {author} {\bibinfo {author} {\bibfnamefont {D.}~\bibnamefont
  {Zhou}}, \bibinfo {author} {\bibfnamefont {A.}~\bibnamefont {Bashan}},
  \bibinfo {author} {\bibfnamefont {Y.}~\bibnamefont {Berezin}}, \bibinfo
  {author} {\bibfnamefont {R.}~\bibnamefont {Cohen}}, \ and\ \bibinfo {author}
  {\bibfnamefont {S.}~\bibnamefont {Havlin}},\ }\href@noop {} {\bibfield
  {journal} {\bibinfo  {journal} {arXiv:1211.2330}\ } (\bibinfo {year}
  {2012})}\BibitemShut {NoStop}%
\bibitem [{\citenamefont {Zhou}\ \emph {et~al.}(2014)\citenamefont {Zhou},
  \citenamefont {Bashan}, \citenamefont {Cohen}, \citenamefont {Berezin},
  \citenamefont {Shnerb},\ and\ \citenamefont {Havlin}}]{zhou2014simultaneous}%
  \BibitemOpen
  \bibfield  {author} {\bibinfo {author} {\bibfnamefont {D.}~\bibnamefont
  {Zhou}}, \bibinfo {author} {\bibfnamefont {A.}~\bibnamefont {Bashan}},
  \bibinfo {author} {\bibfnamefont {R.}~\bibnamefont {Cohen}}, \bibinfo
  {author} {\bibfnamefont {Y.}~\bibnamefont {Berezin}}, \bibinfo {author}
  {\bibfnamefont {N.}~\bibnamefont {Shnerb}}, \ and\ \bibinfo {author}
  {\bibfnamefont {S.}~\bibnamefont {Havlin}},\ }\href@noop {} {\bibfield
  {journal} {\bibinfo  {journal} {Phys. Rev. E}\ }\textbf {\bibinfo {volume}
  {90}},\ \bibinfo {pages} {012803} (\bibinfo {year} {2014})}\BibitemShut
  {NoStop}%
\bibitem [{\citenamefont {Son}\ \emph {et~al.}(2012)\citenamefont {Son},
  \citenamefont {Bizhani}, \citenamefont {Christensen}, \citenamefont
  {Grassberger},\ and\ \citenamefont {Paczuski}}]{son2012percolation}%
  \BibitemOpen
  \bibfield  {author} {\bibinfo {author} {\bibfnamefont {S.-W.}\ \bibnamefont
  {Son}}, \bibinfo {author} {\bibfnamefont {G.}~\bibnamefont {Bizhani}},
  \bibinfo {author} {\bibfnamefont {C.}~\bibnamefont {Christensen}}, \bibinfo
  {author} {\bibfnamefont {P.}~\bibnamefont {Grassberger}}, \ and\ \bibinfo
  {author} {\bibfnamefont {M.}~\bibnamefont {Paczuski}},\ }\href@noop {}
  {\bibfield  {journal} {\bibinfo  {journal} {EPL}\ }\textbf {\bibinfo {volume}
  {97}},\ \bibinfo {pages} {16006} (\bibinfo {year} {2012})}\BibitemShut
  {NoStop}%
\bibitem [{\citenamefont {Dorogovtsev}\ \emph
  {et~al.}(2006{\natexlab{b}})\citenamefont {Dorogovtsev}, \citenamefont
  {Goltsev},\ and\ \citenamefont {Mendes}}]{dorogovtsev2006k}%
  \BibitemOpen
  \bibfield  {author} {\bibinfo {author} {\bibfnamefont {S.~N.}\ \bibnamefont
  {Dorogovtsev}}, \bibinfo {author} {\bibfnamefont {A.~V.}\ \bibnamefont
  {Goltsev}}, \ and\ \bibinfo {author} {\bibfnamefont {J.~F.~F.}\ \bibnamefont
  {Mendes}},\ }\href@noop {} {\bibfield  {journal} {\bibinfo  {journal}
  {Physica D: Nonlin. Phenom.}\ }\textbf {\bibinfo {volume} {224}},\ \bibinfo
  {pages} {7} (\bibinfo {year} {2006}{\natexlab{b}})}\BibitemShut {NoStop}%
\bibitem [{\citenamefont {Birolli}(2007)}]{Birolli2007}%
  \BibitemOpen
  \bibfield  {author} {\bibinfo {author} {\bibfnamefont {G.}~\bibnamefont
  {Birolli}},\ }\href {\doibase doi:10.1038/nphys580} {\bibfield  {journal}
  {\bibinfo  {journal} {Nature Physics}\ }\textbf {\bibinfo {volume} {3}},\
  \bibinfo {pages} {222 } (\bibinfo {year} {2007})}\BibitemShut {NoStop}%
\bibitem [{\citenamefont {Zhou}\ \emph {et~al.}(2013)\citenamefont {Zhou},
  \citenamefont {Gao}, \citenamefont {Stanley},\ and\ \citenamefont
  {Havlin}}]{zhou2013}%
  \BibitemOpen
  \bibfield  {author} {\bibinfo {author} {\bibfnamefont {D.}~\bibnamefont
  {Zhou}}, \bibinfo {author} {\bibfnamefont {J.}~\bibnamefont {Gao}}, \bibinfo
  {author} {\bibfnamefont {H.~E.}\ \bibnamefont {Stanley}}, \ and\ \bibinfo
  {author} {\bibfnamefont {S.}~\bibnamefont {Havlin}},\ }\href {\doibase
  10.1103/PhysRevE.87.052812} {\bibfield  {journal} {\bibinfo  {journal} {Phys.
  Rev. E}\ }\textbf {\bibinfo {volume} {87}},\ \bibinfo {pages} {052812}
  (\bibinfo {year} {2013})}\BibitemShut {NoStop}%
\bibitem [{\citenamefont {Grassberger}(2015)}]{Grassberger2015}%
  \BibitemOpen
  \bibfield  {author} {\bibinfo {author} {\bibfnamefont {P.}~\bibnamefont
  {Grassberger}},\ }\href@noop {} {\bibfield  {journal} {\bibinfo  {journal}
  {arXiv:1502.01623}\ } (\bibinfo {year} {2015})}\BibitemShut {NoStop}%
\bibitem [{\citenamefont {Boccaletti}\ \emph {et~al.}(2014)\citenamefont
  {Boccaletti}, \citenamefont {Bianconi}, \citenamefont {Criado}, \citenamefont
  {Del~Genio}, \citenamefont {G{\'o}mez-Garde{\~n}es}, \citenamefont {Romance},
  \citenamefont {Sendina-Nadal}, \citenamefont {Wang},\ and\ \citenamefont
  {Zanin}}]{boccaletti2014structure}%
  \BibitemOpen
  \bibfield  {author} {\bibinfo {author} {\bibfnamefont {S.}~\bibnamefont
  {Boccaletti}}, \bibinfo {author} {\bibfnamefont {G.}~\bibnamefont
  {Bianconi}}, \bibinfo {author} {\bibfnamefont {R.}~\bibnamefont {Criado}},
  \bibinfo {author} {\bibfnamefont {C.}~\bibnamefont {Del~Genio}}, \bibinfo
  {author} {\bibfnamefont {J.}~\bibnamefont {G{\'o}mez-Garde{\~n}es}}, \bibinfo
  {author} {\bibfnamefont {M.}~\bibnamefont {Romance}}, \bibinfo {author}
  {\bibfnamefont {I.}~\bibnamefont {Sendina-Nadal}}, \bibinfo {author}
  {\bibfnamefont {Z.}~\bibnamefont {Wang}}, \ and\ \bibinfo {author}
  {\bibfnamefont {M.}~\bibnamefont {Zanin}},\ }\href@noop {} {\bibfield
  {journal} {\bibinfo  {journal} {Phys. Reports}\ }\textbf {\bibinfo {volume}
  {544}},\ \bibinfo {pages} {1} (\bibinfo {year} {2014})}\BibitemShut {NoStop}%
\bibitem [{\citenamefont {Kivel{\"a}}\ \emph {et~al.}(2014)\citenamefont
  {Kivel{\"a}}, \citenamefont {Arenas}, \citenamefont {Barthelemy},
  \citenamefont {Gleeson}, \citenamefont {Moreno},\ and\ \citenamefont
  {Porter}}]{kivela2014multilayer}%
  \BibitemOpen
  \bibfield  {author} {\bibinfo {author} {\bibfnamefont {M.}~\bibnamefont
  {Kivel{\"a}}}, \bibinfo {author} {\bibfnamefont {A.}~\bibnamefont {Arenas}},
  \bibinfo {author} {\bibfnamefont {M.}~\bibnamefont {Barthelemy}}, \bibinfo
  {author} {\bibfnamefont {J.~P.}\ \bibnamefont {Gleeson}}, \bibinfo {author}
  {\bibfnamefont {Y.}~\bibnamefont {Moreno}}, \ and\ \bibinfo {author}
  {\bibfnamefont {M.~A.}\ \bibnamefont {Porter}},\ }\href@noop {} {\bibfield
  {journal} {\bibinfo  {journal} {J. Complex Networks}\ }\textbf {\bibinfo
  {volume} {2}},\ \bibinfo {pages} {203} (\bibinfo {year} {2014})}\BibitemShut
  {NoStop}%
\bibitem [{\citenamefont {Baxter}\ \emph {et~al.}(2012)\citenamefont {Baxter},
  \citenamefont {Dorogovtsev}, \citenamefont {Goltsev},\ and\ \citenamefont
  {Mendes}}]{baxter2012avalanche}%
  \BibitemOpen
  \bibfield  {author} {\bibinfo {author} {\bibfnamefont {G.}~\bibnamefont
  {Baxter}}, \bibinfo {author} {\bibfnamefont {S.}~\bibnamefont {Dorogovtsev}},
  \bibinfo {author} {\bibfnamefont {A.}~\bibnamefont {Goltsev}}, \ and\
  \bibinfo {author} {\bibfnamefont {J.}~\bibnamefont {Mendes}},\ }\href@noop {}
  {\bibfield  {journal} {\bibinfo  {journal} {Phys. Rev. Lett.}\ }\textbf
  {\bibinfo {volume} {109}},\ \bibinfo {pages} {248701} (\bibinfo {year}
  {2012})}\BibitemShut {NoStop}%
\bibitem [{\citenamefont {Iwata}\ and\ \citenamefont {Sasa}(2009)}]{iw2009}%
  \BibitemOpen
  \bibfield  {author} {\bibinfo {author} {\bibfnamefont {M.}~\bibnamefont
  {Iwata}}\ and\ \bibinfo {author} {\bibfnamefont {S.-i.}\ \bibnamefont
  {Sasa}},\ }\href {http://stacks.iop.org/1751-8121/42/i=7/a=075005} {\bibfield
   {journal} {\bibinfo  {journal} {J. Phys. A: Math. and Theor.}\ }\textbf
  {\bibinfo {volume} {42}},\ \bibinfo {pages} {075005} (\bibinfo {year}
  {2009})}\BibitemShut {NoStop}%
\bibitem [{\citenamefont {Strogatz}\ and\ \citenamefont
  {Westervelt}(1989)}]{strogatz1989}%
  \BibitemOpen
  \bibfield  {author} {\bibinfo {author} {\bibfnamefont {S.~H.}\ \bibnamefont
  {Strogatz}}\ and\ \bibinfo {author} {\bibfnamefont {R.~M.}\ \bibnamefont
  {Westervelt}},\ }\href@noop {} {\bibfield  {journal} {\bibinfo  {journal}
  {Phys. Rev. B}\ }\textbf {\bibinfo {volume} {40}},\ \bibinfo {pages} {10501}
  (\bibinfo {year} {1989})}\BibitemShut {NoStop}%
\bibitem [{\citenamefont {Strogatz}(1994)}]{Strogatz1994}%
  \BibitemOpen
  \bibfield  {author} {\bibinfo {author} {\bibfnamefont {S.~H.}\ \bibnamefont
  {Strogatz}},\ }\href@noop {} {\emph {\bibinfo {title} {Nonlinear Dynamics and
  Chaos}}}\ (\bibinfo  {publisher} {Perseus Books},\ \bibinfo {address}
  {Massachusetts},\ \bibinfo {year} {1994})\BibitemShut {NoStop}%
\bibitem [{\citenamefont {Goltsev}\ \emph {et~al.}(2006)\citenamefont
  {Goltsev}, \citenamefont {Dorogovtsev},\ and\ \citenamefont
  {Mendes}}]{gdm2006}%
  \BibitemOpen
  \bibfield  {author} {\bibinfo {author} {\bibfnamefont {A.~V.}\ \bibnamefont
  {Goltsev}}, \bibinfo {author} {\bibfnamefont {S.~N.}\ \bibnamefont
  {Dorogovtsev}}, \ and\ \bibinfo {author} {\bibfnamefont {J.~F.~F.}\
  \bibnamefont {Mendes}},\ }\href {\doibase 10.1103/PhysRevE.73.056101}
  {\bibfield  {journal} {\bibinfo  {journal} {Phys. Rev. E}\ }\textbf {\bibinfo
  {volume} {73}},\ \bibinfo {pages} {056101} (\bibinfo {year}
  {2006})}\BibitemShut {NoStop}%
\bibitem [{\citenamefont {Harris}(1989)}]{Harris1989}%
  \BibitemOpen
  \bibfield  {author} {\bibinfo {author} {\bibfnamefont {T.~E.}\ \bibnamefont
  {Harris}},\ }\href@noop {} {\emph {\bibinfo {title} {{The Theory of Branching
  Processes}}}}\ (\bibinfo  {publisher} {Dover},\ \bibinfo {address} {New
  York},\ \bibinfo {year} {1989})\BibitemShut {NoStop}%
\bibitem [{\citenamefont {Zapperi}\ \emph {et~al.}(1995)\citenamefont
  {Zapperi}, \citenamefont {Lauritsen},\ and\ \citenamefont
  {Stanley}}]{zlbs1995}%
  \BibitemOpen
  \bibfield  {author} {\bibinfo {author} {\bibfnamefont {S.}~\bibnamefont
  {Zapperi}}, \bibinfo {author} {\bibfnamefont {K.~B.}\ \bibnamefont
  {Lauritsen}}, \ and\ \bibinfo {author} {\bibfnamefont {H.~E.}\ \bibnamefont
  {Stanley}},\ }\href {\doibase 10.1103/PhysRevLett.75.4071} {\bibfield
  {journal} {\bibinfo  {journal} {Phys. Rev. Lett.}\ }\textbf {\bibinfo
  {volume} {75}},\ \bibinfo {pages} {4071} (\bibinfo {year}
  {1995})}\BibitemShut {NoStop}%
\bibitem [{\citenamefont {Sethna}\ \emph {et~al.}(2001)\citenamefont {Sethna},
  \citenamefont {Dahmen},\ and\ \citenamefont {Myers}}]{sethna2001crackling}%
  \BibitemOpen
  \bibfield  {author} {\bibinfo {author} {\bibfnamefont {J.~P.}\ \bibnamefont
  {Sethna}}, \bibinfo {author} {\bibfnamefont {K.~A.}\ \bibnamefont {Dahmen}},
  \ and\ \bibinfo {author} {\bibfnamefont {C.~R.}\ \bibnamefont {Myers}},\
  }\href@noop {} {\bibfield  {journal} {\bibinfo  {journal} {Nature}\ }\textbf
  {\bibinfo {volume} {410}},\ \bibinfo {pages} {242} (\bibinfo {year}
  {2001})}\BibitemShut {NoStop}%
\bibitem [{\citenamefont {Beggs}\ and\ \citenamefont
  {Plenz}(2003)}]{beggs2003neuronal}%
  \BibitemOpen
  \bibfield  {author} {\bibinfo {author} {\bibfnamefont {J.~M.}\ \bibnamefont
  {Beggs}}\ and\ \bibinfo {author} {\bibfnamefont {D.}~\bibnamefont {Plenz}},\
  }\href@noop {} {\bibfield  {journal} {\bibinfo  {journal} {J. Neurosci.}\
  }\textbf {\bibinfo {volume} {23}},\ \bibinfo {pages} {11167} (\bibinfo {year}
  {2003})}\BibitemShut {NoStop}%
\end{thebibliography}%

\end{document}